\documentclass[twocolumn]{aastex631}

\usepackage{bm}

\usepackage{ulem}
\usepackage{graphicx}

\usepackage{CJK}
\usepackage{soul}
\newcommand{\mathcolorbox}[2]{\colorbox{#1}{$\displaystyle #2$}}

\newcommand{\bmx}{\bm{x}}
\newcommand{\bms}{\bm{s}}
\newcommand{\mr}{\mathrm}
\newcommand{\bmk}{\bm{k}}
\newcommand{\bmq}{\bm{q}}

\begin{document}

\title{Constraining Neutrino Cosmologies with Nonlinear Reconstruction}


\author[0000-0003-2299-6235]{Shi-Hui Zang \begin{CJK*}{UTF8}{gbsn}(臧诗慧)\end{CJK*}}
\affiliation{National Astronomy Observatories, Chinese Academy of Sciences, 20A Datun Road, Beijing 100101, China}

\author[0000-0002-8202-8642]{Hong-Ming Zhu \begin{CJK*}{UTF8}{gbsn}(朱弘明)\end{CJK*}}
\email{hmzhu@nao.cas.cn}
\affiliation{National Astronomy Observatories, Chinese Academy of Sciences, 20A Datun Road, Beijing 100101, China}
\affiliation{Canadian Institute for Theoretical Astrophysics, 60 St. George Street, Toronto, Ontario M5S 3H8, Canada}

\begin{abstract}
Nonlinear gravitational evolution induces strong nonlinearities in the observed cosmological density fields, leading to positive off-diagonal correlations in the power spectrum covariance.
This has caused the information saturation in the power spectrum, e.g., the neutrino mass constraints from the nonlinear power spectra are lower than their linear counterparts by a factor of $\sim2$ at $z=0$.
In this paper, we explore how nonlinear reconstruction methods improve the cosmological information from nonlinear cosmic fields.
By applying nonlinear reconstruction to cold dark matter fields from the Quijote simulations, we find that nonlinear reconstruction can improve the constraints on cosmological parameters significantly, nearly reaching the linear theory limit.
For neutrino mass, the result is only $12\%$ lower than the linear power spectrum, i.e., the theoretical best result.
This makes nonlinear reconstruction an efficient and useful method to extract neutrino information from current and upcoming galaxy surveys.
\end{abstract}


\section{Introduction}

The large-scale structure of the Universe as traced by galaxy clustering provides one of the best windows on fundamental physics including neutrino masses, dark energy, inflation and gravity.
The ongoing and future galaxy redshift surveys have the potential of greatly improving our constraints on cosmology and fundamental physics (e.g., DESI \citealt[][]{2016arXiv161100036D}; Euclid \citealt[][]{2018LRR....21....2A,2020A&A...642A.191E}; SPHEREx \citealt[][]{2014arXiv1412.4872D}; Subaru Prime Focus Spectrograph \citealt[][]{2014PASJ...66R...1T}; Vera Rubin Observatory/LSST \citealt[][]{2009arXiv0912.0201L}; Roman Space Telescope/WFIRST \citealt[][]{2019arXiv190205569A}; MegaMapper \citealt[][]{2022arXiv220904322S}).
To fully exploit these powerful surveys, it is necessary to understand how to optimally extract information from the observed galaxy density fields.

Cosmological constraints derived from galaxy power spectrum are connected to extracting information from linear (or mildly nonlinear) modes, for which the coupling with other modes can be assumed be to subdominant and have evolved  independently from the early stages of the Universe \citep[][]{2022arXiv220307506F}.
However, at low redshifts and on small scales, nonlinear evolution leads to the coupling of modes on different scales and typically causes significant correlations in the small-scale power spectrum
\citep[][]{1999MNRAS.308.1179M,2005MNRAS.360L..82R,2006MNRAS.371.1205R}.
These positive off-diagonal correlations in the power spectrum covariance lead to the information saturation in the power spectrum \citep[][]{2005MNRAS.360L..82R,2006MNRAS.371.1205R,2020ApJS..250....2V,2021ApJ...919...24B}, i.e., the cosmological information does not increase when increasing the wavenumber in the nonlinear regime.
For example, the constraints on neutrino mass from the nonlinear power spectra are lower than their linear counterparts by a factor of 2 at redshift $z=0$ \citep[][]{2022PhRvD.105l3510B}.

The constraints on cosmological parameters including neutrino mass scale approximately linearly with the primordial physics Figure of Merit (FoM), which counts the total number of linear modes accessible to a galaxy redshift survey \citep[see][for more discussions]{2021JCAP...12..049S,2022arXiv220307506F}.
Nonlinear structure formation erases the correlation with the initial conditions and leads to a loss of information, which degrades the FoM of low redshift surveys significantly.
Therefore, recovering more linear modes from the observed nonlinear fields can improve the FoM and cosmological constraints greatly and thus maximize the scientific return of future surveys.



Reconstruction has been an efficient technique to restore the baryonic acoustic oscillation (BAO) feature in the linear power spectrum \citep[][]{2007ApJ...664..675E} and has been widely used to analyse the galaxy survey data (e.g., SDSS \citealt[][]{2012MNRAS.427.2132P}, BOSS \citealt[][]{2017MNRAS.470.2617A}; eBOSS \citealt[][]{2021PhRvD.103h3533A}).
In standard reconstruction, the linear displacement field estimated from the observed nonlinear density under the Zel'dovich approximation is used to move galaxies, which reverses the large-scale bulk flows and partially restores the linear BAO signal \citep[][]{2007ApJ...664..675E}.
Recently nonlinear reconstruction algorithms have been developed to further improve upon the standard reconstruction by solving the full nonlinear displacement field \citep[][]{2017PhRvD..96l3502Z,2017PhRvD..96b3505S, 2018PhRvD..97b3505S}.
The reconstruction methods works extraordinarily well for reducing the uncertainty on the BAO scale. 
Slightly more generally, the reconstruction can restore information about the linear modes that were corrupted by nonlinear structure formation and a higher primordial FoM can be obtained.
It has been shown that nonlinear reconstruction reduces the off-diagonal components of power spectrum covariance prominently and increase the information content of power spectrum amplitude substantially \citep[see e.g.,][]{2017MNRAS.469.1968P}. Recently, \citet{2022arXiv220205248W} have explored the improvement of cosmological constraints using the standard reconstruction algorithm, while \citet{2023arXiv230507018F} have studied improving constraints using neural network based reconstruction, in particular primordial non-Gaussianity.

In this paper, we explore how nonlinear reconstruction improves the constraints on cosmological parameters, with a focus on neutrino mass.
This improves upon the previous study \citep[][]{2017MNRAS.469.1968P}, which only considered one parameter, the power spectrum amplitude.
We use the Quijote simulations \citep{2020ApJS..250....2V} to compute the pre- and post-reconstruction power spectrum covariance and perform a detailed Fisher analysis.
We find that nonlinear reconstruction can efficiently reduce the mode-coupling effects in the nonlinear field and improve the cosmological constraints significantly, almost reaching the linear theory limit, which is an important step towards optimal extraction of neutrino mass information on small scales and has profound implications for future galaxy redshift surveys.

This paper is organized as follows. 
In Section~\ref{sec:Reconstruction}, we describe the reconstruction algorithms and Fisher matrix. 
In Section~\ref{sec:Simulations}, we present the numerical implementation of the Fisher analysis. 
In Section~\ref{sec:Results}, we show the numerical results. 
We discuss and conclude in Section~\ref{sec:Discussion}.

\section{Method}
\label{sec:Reconstruction}

Neutrinos with finite mass suppress the small-scale power spectrum below the neutrino free-streaming scale \citep[][]{1980PhysRevLett.45.1980,2006PhR...429..307L}.
The suppression is proportional to the sum of neutrino masses, thus allowing us to measure or constrain the total neutrino mass $M_\nu$.
Detecting the minimum neutrino mass from the effects of neutrinos on the total matter $\delta_m$ and cold dark matter (CDM)+baryon perturbations $\delta_{cb}$ is one of the key science goals of the near-term and future galaxy surveys.

For galaxy clustering in the presence of massive neutrinos, the now-standard approximation that galaxies trace the CDM+baryon density field $\delta_{cb}$ has been shown to be an excellent approximation well into the quasi-linear regime \citep{2015JCAP...07..043C}.
The galaxy overdensity is given by $\delta_g=F[\delta_{cb}]$ and at linear order $\delta_g=b\delta_{cb}$, where $b$ is the linear galaxy bias.
Recently, \citet[][]{2022PhRvD.105l3510B} demonstrated that all the information of neutrinos in the CDM field is in the linear power spectrum set by the initial conditions for $k\lesssim1\ h\mathrm{Mpc}^{-1}$ using the phase-matched $N$-body simulations.
Thus, galaxy clustering, which is based on the $\delta_{cb}$ field will have negligible constraining power on neutrino masses beyond that which exists in the linear density field over the same range of scales \citep[][]{2022PhRvD.105l3510B}.
They have found similar conclusions for other parameters such as the total matter density $\Omega_m$ and primordial power spectrum amplitude $A_s$.
While the late-time linear power spectrum is not an observable, it provides a bound on the error of neutrino masses available from galaxy clustering.
However, the nonlinear power spectrum does not reach this bound, because of the positive off-diagonal components in its covariance which degrades the constraints substantially.
The resulting constraining power of the nonlinear power spectrum for $\delta_{cb}$ is a factor of 2 lower than the linear one \citep[][]{2022PhRvD.105l3510B}.

Therefore, there is still room for improving neutrino mass constraints from galaxy clustering and potentially a factor of 2 better constraints can be achieved.
The goal of this paper is to investigate whether nonlinear reconstruction can recover the information from the nonlinear $\delta_{cb}$ field to the linear theory limit. 

In this section, we review the reconstruction algorithms including standard and nonlinear reconstruction and describe the Fisher analysis of cosmological information.

\subsection{Reconstruction}

Reconstruction is an efficient way to reduce nonlinearities induced by the large-scale bulk flows.
The standard reconstruction has been first proposed to reconstruct the linear BAO signal \citep[][]{2007ApJ...664..675E}.
The algorithm devised by \citet[][]{2007ApJ...664..675E} consists of the following steps:
\begin{itemize}
    \item Smooth the nonlinear density field with a Gaussian kernel $W_R(k)$ of scale $R$, to filter out small-scale modes.
    \item Compute the shift $\bm{s}$ from the smoothed density field using the Zel'dovich approximation,
    \begin{equation}
    \bm{s}(\bm{k}) = -\frac{ik}{k^2}W_R(k)\delta(\bm{k}).
    \end{equation}
    \item Move the galaxies by $\bm{s}$ to compute the ``displaced'' density field, $\delta_d(\bmx)$.
    \item Shift an spatially uniform distribution of particles by $\bms$ to form the shifted density field, $\delta_s(\bmx)$.
    \item The reconstructed density field is defined as $\delta_{r}(\bm{x})\equiv\delta_d(\bm{x})-\delta_s(\bm{x})$.
\end{itemize}

The Zel'dovich approximation only captures the linear displacement on large scales, limiting the efficacy of standard reconstruction to relatively large scales, $k\sim0.2\ h\mr{Mpc}^{-1}$.
Recent new reconstruction algorithms have been presented to solve the nonlinear displacement field \citep[][]{2017PhRvD..96l3502Z,2017PhRvD..96b3505S,2018PhRvD..97b3505S}, which improve upon the linear standard reconstruction with a substantially improved small-scale performance. 
While these methods are different in the detailed numerical implementations, the underlying physical principles are the same, i.e., solving the full displacement from the mass conservation relation. 
In the following analysis, we use the algorithm of \citet[][]{2017PhRvD..96b3505S}, solving the displacement field iteratively using the fast Fourier transform, which is efficient and easy to apply.
The reconstruction algorithm devised by \citet[][]{2017PhRvD..96b3505S} consists of the following steps: 
\begin{itemize}
    \item Iteratively displace objects as follows:
    \begin{itemize}
        \item[1.] From the catalog, compute the density field on the grid.
        \item[2.] Smooth the density field with the Gaussian window $W_R(k)$, with smoothing scale 
        \begin{equation}
        R=\mathrm{max}(\epsilon_R^{n-1}R_{\mathrm{init}}, R_{\mathrm{min}}),
        \end{equation}
        in the $n$th iteration step and truncate small-scale modes by setting $\delta(\bmk)=0$ if $k>k_{\mr{Ny}}$, where $k_{\mr{Ny}}$ is the Nyquist frequency.
        \item[3.] Compute the negative Zel'dovich displacement
        \begin{equation}
        \bm{s}(\bm{k}) = -\frac{ik}{k^2}W_R(k)\delta(\bm{k}).
        \end{equation}
        \item[4.] Move the object by the displacement $\bms$ and update the object's position. 
    \end{itemize}
    \item Compute the total displacement of each object $\bm{\chi}(\hat{\bm{q}}_{\mathrm{end}})=\hat{\bm{q}}_{\mathrm{end}}-\bm{x}_{\mathrm{start}}$, where $\bm{x}_{\mathrm{start}}$ is the starting position before the first iteration step and $\hat{\bm{q}}_{\mathrm{end}}$ is the end position of each object after iteration ($\bmx_{\mr{start}}$ is the Eulerian position of each object and $\hat{\bm{q}}_{\mathrm{end}}$ is the estimated Lagrangian position).
    \item Paint the estimated displacement $\bm{\chi}(\hat{\bm{q}}_{\mathrm{end}})$  defined on the estimated Lagrangian position  $\hat{\bm{q}}_{\mathrm{end}}$ to a regular grid and truncate the small-scale modes beyond $k_{\mr{Ny}}$.
    \item The reconstructed density field from nonlinear reconstruction is 
    \begin{equation}
        \delta_{nr}(\bmq)=\nabla_q \cdot \bm{\chi}(\bmq).
    \end{equation}
\end{itemize}

An improved estimation of the small-scale displacement allows us to recover more small-scale linear modes.
The cross-correlation coefficient of the reconstructed field with the linear initial conditions is higher than 0.5 for $k\lesssim1\ h\mr{Mpc}^{-1}$ \citep[][]{2017PhRvD..96l3502Z,2017PhRvD..96b3505S,2018PhRvD..97b3505S}.
Note that such an improvement can also be achieved by combining the standard reconstruction and  convolutional neural networks \citep[see][]{2022arXiv220712511S,2023arXiv230507018F}.

\subsection{Fisher analysis}

To quantify the cosmological information in the power spectrum of pre- and post-reconstruction fields, we use the Fisher information matrix \citep[][]{1925PCPS...22..700F}.
The Fisher matrix is defined as
\begin{equation}
    F_{\alpha\beta}=\sum_{i,j}\frac{\partial P_i}{\partial \theta_\alpha}C_{ij}^{-1}\frac{\partial P_j}{\partial \theta_\beta},
\end{equation}
where $P_i=P(k_i)$ is the power spectrum of the $i$th bin and $C_{ij}$ is the power spectrum covariance matrix, and $\theta_a$ represents the cosmological parameters \citep[][]{1997PhRvL..79.3806T}.
We have neglected the dependence of covariance on the cosmological parameters to avoid underestimating the errors \citep[][]{2013A&A...551A..88C}.

The inverse of the Fisher matrix gives the covariance matrix of the parameters and the expected errors on the parameters are equal to the square root of the diagonal elements of $F^{-1}$,
\begin{equation}
    \sigma(\theta_\alpha)= \sqrt{(F^{-1})_{\alpha\alpha}},
\end{equation}
which represents the minimum error achievable.

\section{Simulations}
\label{sec:Simulations}

To compute the covariance matrix of the power spectrum and the response of power spectrum with respect to the cosmological parameters, we use the Quijote simulations \citep[][]{2020ApJS..250....2V}, run with the TreePM code {\tt Gadget-III}, an improved version of {\tt Gadget-II} \citep[][]{2005MNRAS.364.1105S}.
Each simulation evolves $512^3$ CDM particles (plus $512^3$ neutrino particles for nonzero neutrino mass) in a cosmological volume of $1\ (h^{-1}\mr{Gpc})^3$.
The Quijote simulations have been widely used to quantify the information content of nonlinear statistics \citep[e.g.,][]{2020ApJS..250....2V,2020JCAP...03..040H,2021PhRvL.126a1301M,2021ApJ...919...24B,2022arXiv220904310P}.
More details of the Quijote simulations can be found in \citet[][]{2020ApJS..250....2V}.

We consider six cosmological parameters, $\Omega_m$, $\Omega_b$, $h$, $n_s$, $\sigma_8$, and $M_\nu$.
We use $5,000$ simulations of the fiducial cosmology to compute the covariance matrix. 
We use $1,000$ simulations to compute the derivatives with respect to each parameter.
The detailed information of simulations we used in the analysis is shown in Table~\ref{table:Quijote}.
\begin{table*}[tbh!]
\begin{tabular}{ccccccccc}
\hline
Name     & $\Omega_m$                          & $\Omega_b$                         & $h$                                 & $n_s$                               & $\sigma_8$                         & $M_{\nu} \ [\mathrm{eV}]$                    & ICs        & Realizations \\ \hline
Fiducial        & 0.3175                              & 0.049                              & 0.6711                              & 0.9624                              & 0.834                              & 0.0                              & 2LPT       & 5,000         \\
Fiducial ZA     & 0.3175                              & 0.049                              & 0.6711                              & 0.9624                              & 0.834                              & 0.0                              & Zel'dovich & 500          \\
$\Omega_m^+$    & \underline{0.3275} & 0.049                              & 0.6711                              & 0.9624                              & 0.834                              & 0.0                              & 2LPT       & 500          \\
$\Omega_m^-$    & \underline{0.3075} & 0.049                              & 0.6711                              & 0.9624                              & 0.834                              & 0.0                              & 2LPT       & 500          \\
$\Omega_b^{++}$ & 0.3175                              & \underline{0.051} & 0.6711                              & 0.9624                              & 0.834                              & 0.0                              & 2LPT       & 500          \\
$\Omega_b^{--}$ & 0.3175                              & \underline{0.047} & 0.6711                              & 0.9624                              & 0.834                              & 0.0                              & 2LPT       & 500          \\
$h^+$           & 0.3175                              & 0.049                              & \underline{0.6911} & 0.9624                              & 0.834                              & 0.0                              & 2LPT       & 500          \\
$h^-$           & 0.3175                              & 0.049                              & \underline{0.6511} & 0.9624                              & 0.834                              & 0.0                              & 2LPT       & 500          \\
$n_s^+$         & 0.3175                              & 0.049                              & 0.6711                              & \underline{0.9824} & 0.834                              & 0.0                              & 2LPT       & 500          \\
$n_s^-$         & 0.3175                              & 0.049                              & 0.6711                              & \underline{0.9424} & 0.834                              & 0.0                              & 2LPT       & 500          \\
$\sigma_8^+$    & 0.3175                              & 0.049                              & 0.6711                              & 0.9624                              & \underline{0.849} & 0.0                              & 2LPT       & 500          \\
$\sigma_8^-$    & 0.3175                              & 0.049                              & 0.6711                              & 0.9624                              & \underline{0.819} & 0.0                              & 2LPT       & 500          \\
$M_{\nu}^+$     & 0.3175                              & 0.049                              & 0.6711                              & 0.9624                              & 0.834                              & \underline{0.1} & Zel'dovich & 500          \\  \hline
\end{tabular}
\caption{The details of the Quijote simulations used in this work. 
The fiducial cosmology contains $5,000$ simulations, which are used to estimate the power spectrum covariance matrix. 
Other simulations with different cosmological parameters are used to compute the partial derivatives of the power spectrum to cosmological parameters, where only the considered parameter is varied and other
parameters are fixed.
The initial conditions are generated at $z=127$ using the second-order Lagrangian perturbation theory, except for the neutrino simulations where the Zel'dovich approximation is used.
\label{table:Quijote}}
\end{table*}
We have performed a detailed convergence test and verified that our results are converged, i.e., the constraints do not change much if the power spectrum covariance and derivatives are evaluated with less simulations (See Appendix~\ref{sec:Convergence} for more details).

We use the snapshot of CDM particles from the simulation output at redshift $z=0$.
Here we assume baryons trace CDM.
We compute the nonlinear $\delta_{cb}$ field on a $512^3$ grid using the cloud-in-cell interpolation scheme.
For standard reconstruction, we use a Gaussian smoothing of scale $R=10\ h^{-1}\mathrm{Mpc}$, $W_R(k)=e^{-(kR)^2/2}$ following \citet[][]{2007ApJ...664..675E}.
For nonlinear reconstruction, we use $R_{\mathrm{init}}=10\ h^{-1}\mathrm{Mpc}$, $\epsilon_R=0.5$, and $R_{\mathrm{min}} = 1\ h^{-1}\mathrm{Mpc}$.
We use eight iteration steps in the reconstruction \citep[][]{2017PhRvD..96b3505S}.

\subsection{Covariance matrix}

We estimate the covariance matrix using $5,000$ simulations of the fiducial cosmology as
\begin{equation}
    \hat{C}_{ij} = \langle (P_i - \langle P_i\rangle)(P_j - \langle P_j\rangle) \rangle,
\end{equation}
where $\langle \cdot \rangle$ denotes the ensemble average over simulations.
The covariance matrix for fields $\delta_{cb}$, $\delta_r$ and $\delta_{nr}$ are estimated numerically using the Quijote simulations.

For the linear density field (a Gaussian random field), the power spectrum covariance matrix is diagonal,
\begin{equation}
    C_{ij} = \frac{2[P(k_i)]^2}{N_{k_i}}\delta_{ij},
\end{equation}
where $N_{k_i}$ is the number of Fourier modes in the corresponding $k_i$ bin.
This can be directly computed using the linear power spectrum at $z=0$ from the linear Boltzmann code CAMB \citep[][]{2000ApJ...538..473L}.

Since the covariance matrix is estimated from simulations, it is itself a random variable with errors.
The matrix inversions lead to biased constraints on the parameters.
To account for this, we apply a correction to the covariance matrix \citep[][]{2022MNRAS.510.3207P,2022arXiv220904310P}
\begin{equation}
    C = \frac{n_{\mathrm{sim}} - 1}{n_{\mathrm{sim}} - n_{\mathrm{bins}} + n_{\theta} - 1}\hat{C},
\end{equation}
where $n_{\mathrm{sim}}$ is the number of simulations to calculate the covariance matrix, $n_{\theta}$ is the number of parameters in Fisher analysis, $n_{\mathrm{bins}}$ is the number of power spectrum bins. 
In our case, we have $5,000$, $n_{\theta} = 6$, and $n_{\mathrm{bins}}$ depends on the maximum wavenumber $k_{\mathrm{max}}$ included in the analysis.
We have checked that this correction factor is less than $1\%$ for $k_{\mathrm{max}}<0.5\ h\mathrm{Mpc}^{-1}$.

\subsection{Derivatives}

For the parameters $\Omega_m, \Omega_b, h, n_s$ and $\sigma_8$, we use a central difference scheme to compute the derivatives,
\begin{equation}
\label{eq:derivative}
    \frac{\partial P(k)}{\partial \theta_i} \simeq \frac{P(k)|_ {\theta_i + \delta\theta_i} - P(k)|_{\theta_i - \delta\theta_i}}{2\delta\theta_i},
\end{equation}
where $P(k)$ is the measured power spectrum from simulations and only the considered parameter is varied and other parameters are fixed.
We have used 500 pairs of simulations to calculate each derivative using Eq.~(\ref{eq:derivative}). 

For neutrino mass $M_\nu$, the second term in the numerator of Eq.~(\ref{eq:derivative}) corresponds to a cosmology with negative neutrino mass, since the fiducial value of the neutrino mass is at the boundary of the allowed region, i.e., $M_\nu=0$.
Therefore, we use a first-order forward difference scheme to compute the derivative,
\begin{eqnarray}
    \label{eq:derivative_2}
    \frac{\partial P(k)}{\partial M_{\nu}} \simeq && \frac{1}{\delta M_{\nu }} \Big[ P(k)|_{M_\nu+\delta M_\nu}-P(k)|_{M_\nu} \Big],
\end{eqnarray}
where the fiducial neutrino mass $M_\nu=0$.
We use 1,000 simulations to compute the derivative with respect to the neutrino mass.

\section{Results}
\label{sec:Results}

In this section, we present the reconstruction results, including the cross-correlation coefficient, power spectrum covariance matrix, and cosmological parameter constraints.

\subsection{Cross-correlation coefficient}



To quantify the reconstruction performance, we use the cross-correlation coefficient,
\begin{equation}
    r_{\delta_A\delta_B}(k) = \frac{P_{\delta_A\delta_B}(k)}{\sqrt{P_{\delta_A\delta_A}(k)P_{\delta_B\delta_B}(k)}},
\end{equation}
where $P_{\delta_A\delta_A}(k)$ and $P_{\delta_B\delta_B}(k)$ are the power spectrum of fields $\delta_A(k)$ and $\delta_B(k)$, and $P_{\delta_A\delta_B}(k)$ is the cross-power spectrum between the two fields.

\begin{figure}[ht!]
\centering
\includegraphics[width=\columnwidth]{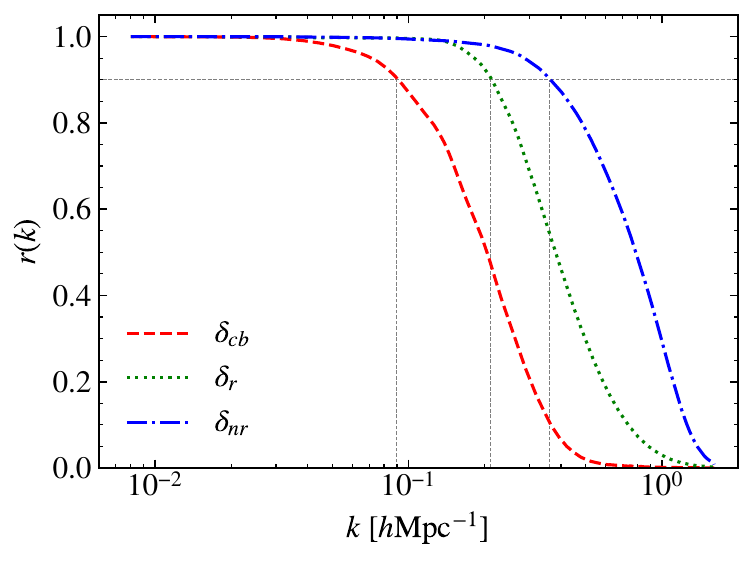}
\caption{The cross-correlation coefficient with the linear initial conditions for the nonlinear $\delta_{cb}$ field, the reconstructed fields using standard reconstruction $\delta_r$ and nonlinear reconstruction $\delta_{nr}$.
While both reconstruction algorithms improve the correlation with the linear initial conditions substantially, the nonlinear method has a better performance on smaller scales, $k\gtrsim0.2\ h\mr{Mpc}^{-1}$.
\label{fig:Power_Corr}}
\end{figure}

In Figure~\ref{fig:Power_Corr}, 
we plot the cross-correlation coefficient with the linear initial conditions $\delta_L$ for the nonlinear $\delta_{cb}$ field, the reconstructed fields using standard reconstruction $\delta_r$ and nonlinear reconstruction $\delta_{nr}$.
We note that the both reconstruction algorithms improve the correlation with the linear density field substantially, especially at small scales.
While being competitive with nonlinear reconstruction at large scales $k\lesssim0.2\ h\mr{Mpc}^{-1}$, the standard reconstruction has a lower correlation with the linear field $\delta_L$ at smaller scales $k\gtrsim0.2\ h\mr{Mpc}^{-1}$, which needs to be improved by solving the small-scale displacement or using the convolutional neural network 
\citep[e.g.][]{2017PhRvD..96l3502Z,2017PhRvD..96b3505S,2018PhRvD..97b3505S,2022arXiv220712511S}.
Nevertheless, we expect that both reconstruction methods can recover more small-scale linear modes, i.e., a much higher primordial physics FoM, and therefore improve the cosmological constraints. 

Note that reconstruction algorithms based on solving the full nonlinear displacement have a similar performance with correlation coefficient  $r(k)\gtrsim0.5$ at $k\lesssim1\ h\mr{Mpc}^{-1}$ \citep[][]{2017PhRvD..96l3502Z,2017PhRvD..96b3505S,2018PhRvD..97b3505S}.
To reduce the computational cost, we have used the iterative algorithm of \citet[][]{2017PhRvD..96b3505S}.

\subsection{Full shape reconstructed power spectrum}
\begin{figure}[ht!]
\centering
\includegraphics[width=\columnwidth]{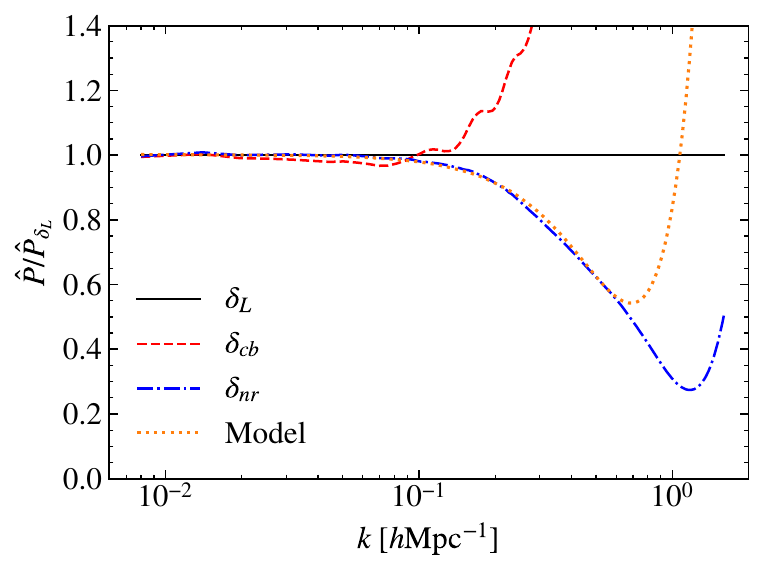}
\caption{Power spectra from the simulations at $z=0$, divided by the linear initial power spectrum linearly scaled to $z=0$. 
The reconstructed power spectrum agrees with the linear power spectrum within $5\%$
at $k<0.16\ h\mr{Mpc}^{-1}$, while the original nonlinear power and linear power agree within $5\%$ at $k<0.15\ h\mr{Mpc}^{-1}$. 
Compared to the nonlinear density without reconstruction, nonlinear reconstruction slightly improves the agreement with the linear linear power spectrum.
However, the best-fit theoretical model agrees with the reconstructed power spectrum within $1\%$ at $k\lesssim0.24\ h\mathrm{Mpc}^{-1}$ and $3\%$ at $k\lesssim0.5\ h\mathrm{Mpc}^{-1}$.}
\label{fig:Power_Ratio}
\end{figure}

Figure~\ref{fig:Power_Ratio}
shows the measured power spectra divided by the linear initial power spectrum linearly scaled to the redshift $z=0$ of the simulation.
The original nonlinear power and linear power agree within $5\%$ at $k<0.15\ h\mr{Mpc}^{-1}$, and the reconstructed power spectrum agrees with the linear power spectrum within $5\%$
at $k<0.16\ h\mr{Mpc}^{-1}$.
Nonlinear reconstruction slightly extends the $k$ range where the linear theory is valid. 
As we have noticed in Fig.~\ref{fig:Power_Corr}, the cross-correlation coefficient $r$ becomes smaller than unity on smaller scales, we would have to add corrections to the linear power spectrum to model the power spectrum after reconstruction.

The fact that nonlinear reconstruction misses small-scale broadband power relative to the linear power is not necessarily a problem.
All that is needed to estimate cosmological parameters is a theoretical model describing the reconstructed power spectrum as a function of cosmological parameters.
Following the literature on modeling the reconstructed power spectrum \citep[see e.g.][]{2009PhRvD..80l3501N,2009PhRvD..79f3523P,2015MNRAS.450.3822W},
we use a simple theoretical model to fit the power spectrum after reconstruction,
\begin{equation}
    \label{eq:pk_theory}
    P_{\mathrm{theory}}(k) = C^2(k)P_{\delta_L}(k) + P_{\mr{N}},
\end{equation}
where the propagator $C(k)=e^{-k^2\Sigma^2/2}$ describes the damping of the linear power spectrum, $P_{\delta_L}$ is the linear initial power spectrum linearly scaled to redshift $z=0$, and the noise power spectrum $P_{\mr{N}}$ quantifies the nonlinearities after reconstruction.
We then fit the two parameters $\Sigma$ and $P_{\mr{N}}$ by minimizing the sum of squares
\begin{equation}
    S=\sum_i\left(\hat{P}_{\delta_{nr}}(k_i)-P_{\mr{theory}}(k_i)\right)^2,
\end{equation}
where the hat denotes the reconstructed power spectrum estimated from simulations and $k_i$ denotes the power spectrum in the $i$th bin. 
Here, we directly use the measured linear initial power spectrum $\hat{P}_{\delta_L}$ when computing the theoretical model, to reduce the impact of variations due to the cosmic variance.
The maximum wavenumber included in the fitting is $k_{\mr{max}}=0.8\ h\mr{Mpc}^{-1}$.
We use equal weights for all $k$ bins to avoid over-fitting small scales.
The best-fit parameters are $\Sigma = 1.67\ h^{-1}\mathrm{Mpc}$ and $P_{\mr{N}}=38 \ [(h^{-1}\mathrm{Mpc})^3]$.
The best-fit theoretical model agrees with the reconstructed power spectrum within $1\%$ at $k\lesssim0.24\ h\mathrm{Mpc}^{-1}$ and $3\%$ at $k\lesssim0.5\ h\mathrm{Mpc}^{-1}$, which makes it possible to perform cosmological parameter inference at the percent level using the reconstructed power spectrum.
However, when applying to galaxy surveys, more complexities have to be considered such as galaxy biasing and redshift space distortions, which will be explored in the future.

On smaller scales, $k\gtrsim0.5\ h\mathrm{Mpc}^{-1}$, the model predicts a higher noise level than the data.
This is partly due to that the simple white noise term misses the scale-dependence of the residual nonlinearities.
Additional improvements on small scales may be possible by using the effective field theory of large-scale structure.
On small scales we would have to add more counterterms to this simple model to match the reconstructed power spectrum \citep[e.g.,][]{2017PhRvD..96b3505S}.
We leave this for future work.

In the following analysis, we use the reconstructed power spectrum in the simulation to compute the Fisher information.

\subsection{Covariance matrix}

We apply two reconstruction algorithms to the $5,000$ simulations of fiducial cosmology and compute the covariance matrix of the power spectrum.

\begin{figure*}[ht!]
    \centering
    \includegraphics[width=\textwidth]{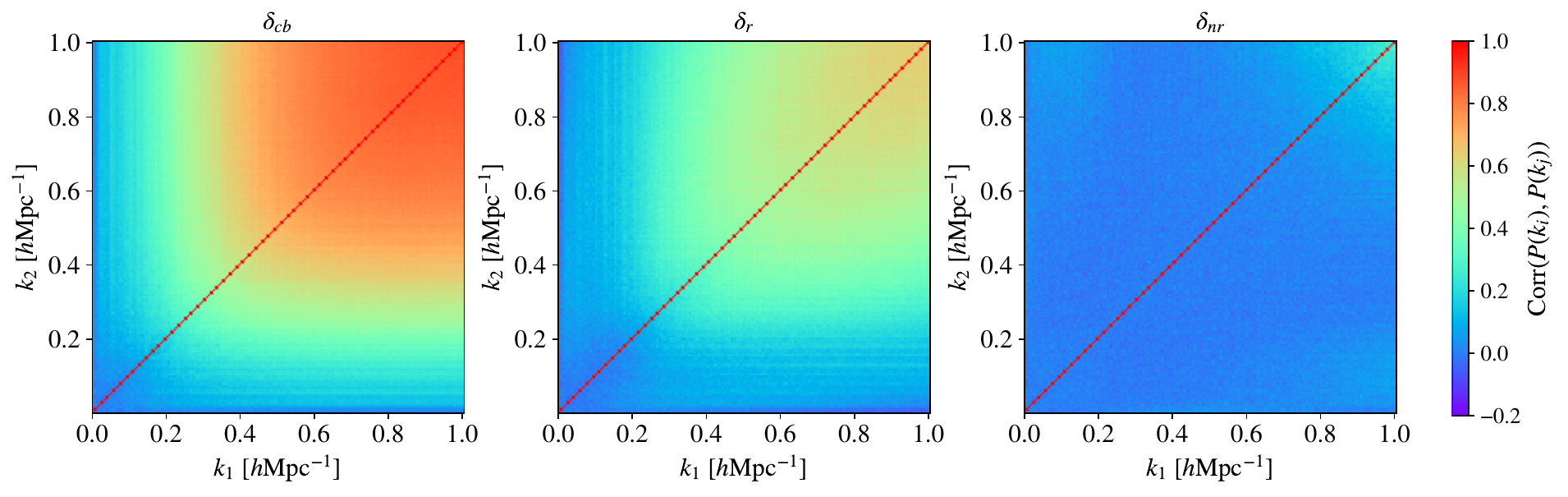}
    \caption{Correlation matrix for the power spectrum of the nonlinear $\delta_{cb}$ field, and the reconstruction fields $\delta_r$ and $\delta_{nr}$ using standard and nonlinear reconstruction methods, respectively.
    The covariance matrix is estimated using $5,000$ simulations.
    Reconstruction reduces the off-diagonal correlation substantially.
    The density modes are nearly independent at $k\lesssim0.8\ h\mr{Mpc}^{-1}$ for the nonlinear reconstruction method.
    \label{fig:Cov}}
\end{figure*}

In Figure~\ref{fig:Cov}, 
we plot the correlation matrix, i.e., the normalized covariance matrix, defined as
$C_{ij}/\sqrt{C_{ii}C_{jj}}$, where $C_{ij}$ is the covariance matrix.
For the nonlinear $\delta_{cb}$ field, the correlation matrix is almost diagonal on large scales, $k\lesssim0.1\ h\mr{Mpc}^{-1}$, and exhibits significant off-diagonal correlations on smaller scales, $k\gtrsim0.1\ h\mr{Mpc}^{-1}$, which leads to the information saturation of the nonlinear power spectrum \citep[e.g.,][]{2005MNRAS.360L..82R,2006MNRAS.371.1205R,2020ApJS..250....2V,2021ApJ...919...24B}.
The standard reconstruction reduces the off-diagonal elements and the covariance matrix is nearly diagonal at $k\lesssim0.2\ h\mr{Mpc}^{-1}$, in agreement with the cross-correlation coefficient observed in Figure~\ref{fig:Power_Corr}.
However, the covariance matrix of nonlinear reconstruction is diagonal well into the nonlinear regime, $k\lesssim0.8\ h\mr{Mpc}^{-1}$, where the coupling with other modes are subdominant.
This nearly diagonal covariance matrix for the $\delta_{nr}$ field indicates more linear information available than the power spectra of the $\delta_{cb}$ and $\delta_r$ fields. 

With enough simulations, we have not observed large negative off-diagonal elements before and after reconstruction.
There only exists a few negative off-diagonal elements with very small absolute values $\sim10^{-2}$, mainly due to the random statistical fluctuations.
Therefore, our results should be more robust than the previous work which uses less realizations to estimate the covariance matrix \citep[][]{2017MNRAS.469.1968P}.
We have confirmed that our results have converged, i.e., the parameter constraints do not change much if less simulations are used.
See Appendix~\ref{sec:Convergence} for more details of the convergence tests.

\subsection{Cosmological constraints}

\begin{figure*}[ht!]
    \centering
    \includegraphics[width=\textwidth]{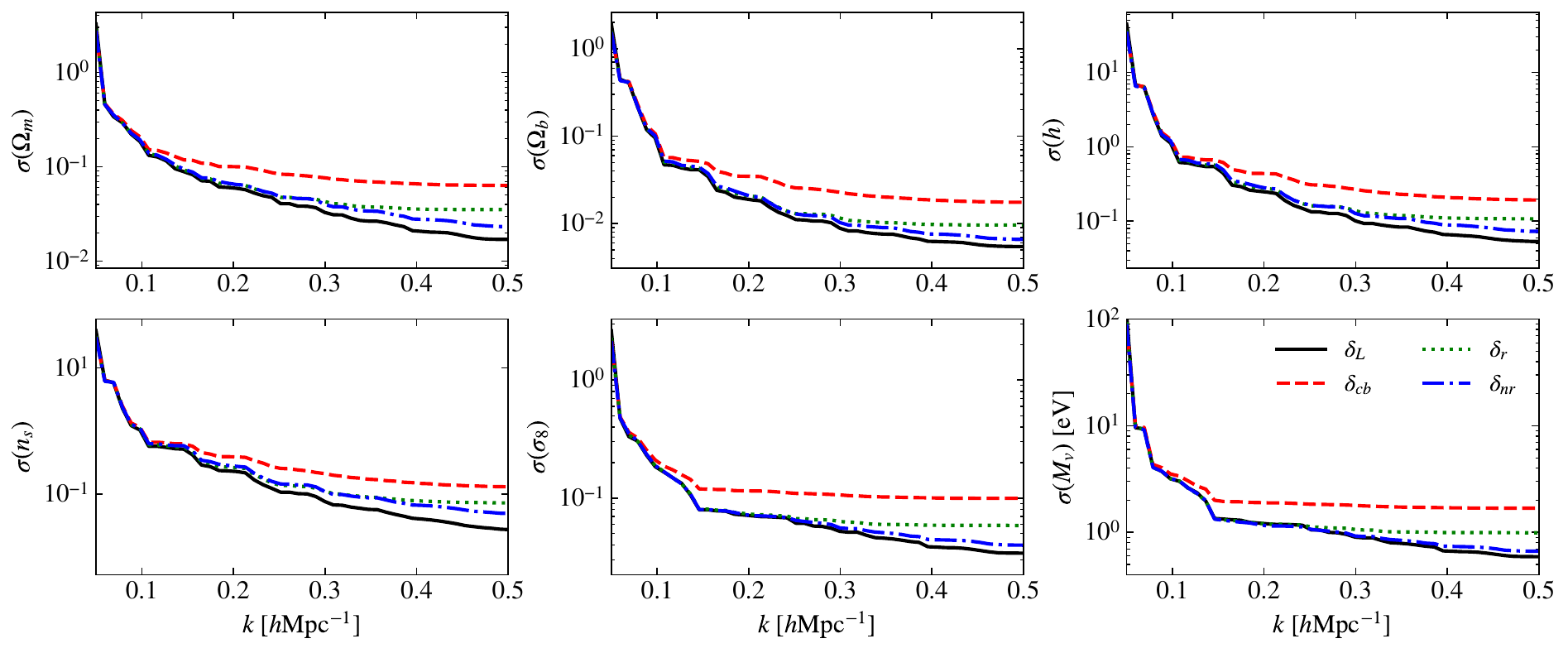}
    \caption{The marginalized $1\sigma$ error for each cosmological parameters as a function of $k_{\mathrm{max}}$.
    The parameter constraints from the nonlinear $\delta_{cb}$ start to saturate at $k_{\mr{max}}=0.2\ h\mr{Mpc}^{-1}$.
    The information from standard reconstruction continues to increase until $k_{\mr{max}}\sim0.3\ h\mr{Mpc}^{-1}$, limited by the linear assumption when estimating the displacement field.
    The constraining power of the $\delta_{nr}$ field is much better than $\delta_{cb}$ and $\delta_r$ and continues to improve until $k_{\mr{max}}\sim0.5\ h\mr{Mpc}^{-1}$.
\label{fig:Constraint}}
\end{figure*}

In Figure~\ref{fig:Constraint}, 
we show the marginalized $1\sigma$ constraints on cosmological parameters, $\Omega_m$, $\Omega_b$, $h$, $n_s$, $\sigma_8$, and $M_\nu$, as a function of $k_{\mr{max}}$.
On large scales where the cosmic density field evolved linearly from the initial conditions, all fields, $\delta_L$, $\delta_{cb}$, $\delta_r$, and $\delta_{nr}$, have a similar constraint on the parameters at $k\lesssim0.1\ h\mr{Mpc}^{-1}$.
The constraints from the nonlinear $\delta_{cb}$ start to saturate at $k_{\mr{max}}=0.2\ h\mr{Mpc}^{-1}$, especially for $\sigma_8$ and $M_\nu$, for which the dominant information comes from the power spectrum amplitude.
For standard reconstruction, the constraints are much better than the nonlinear $\delta_{cb}$ field.
The information from $\delta_r$ continue to increase for $k\gtrsim0.2\ h\mr{Mpc}^{-1}$ and saturates at $k_{\mr{max}}\sim0.3\ h\mr{Mpc}^{-1}$.
This is due to the limited validity of the Zel'dovich approximation adopted in the standard reconstruction. 
However, the constraining power of the $\delta_{nr}$ field is even better than $\delta_r$ and the constraints obtained using nonlinear reconstruction continue to improve until $k_{\mr{max}}=0.5\ h\mr{Mpc}^{-1}$, nearly reaching the linear theory constraints.
The improvement at $k\gtrsim0.3\ h\mr{Mpc}^{-1}$ upon the standard reconstruction is mainly because a better estimation of small-scale displacement using nonlinear methods removes more shift nonlinearities, which is not captured by the Zel'dovich approximation.

We notice that the constraints on neutrino mass $M_\nu$ with reconstruction are almost the same as the linear field at $k\lesssim0.25\ h\mr{Mpc}^{-1}$ and the constraint still nearly reaches the linear theory limit at $k\gtrsim0.25\ h\mr{Mpc}^{-1}$ for the nonlinear reconstruction method.
This result is remarkable, since the $\delta_{cb}$ field and derived quantities, e.g., the halo field, contain little information about neutrino mass beyond that which exists in the linear power spectrum for $k\lesssim1\ h\mr{Mpc}^{-1}$, demonstrated by using phase-matched simulations \citep[][]{2022PhRvD.105l3510B}.
Therefore, the linear power spectrum provides a bound on the error that can be obtained using $\delta_{cb}$ and it is exciting to see that nonlinear reconstruction can recover nearly all the information regarding neutrino mass from the nonlinear $\delta_{cb}$ field.

The $M_\nu-\sigma_8$ degeneracy has limited the neutrino mass constraint on small scales \citep[][]{2020ApJS..250....2V,2021ApJ...919...24B},
which could not be reduced using power spectrum alone 
(See Figure~\ref{fig:Contour} of Appendix~\ref{sec:confidence}).
Combining other nonlinear statistics such as halo mass function and void size function can tighten the constraint on neutrino mass by breaking the degeneracy between different parameters \citep[e.g.][]{2021ApJ...919...24B}, which will be subject of further inquiry.

In Table~\ref{tab:Constraint}, we list the marginalized $1\sigma$ errors for $k_{\mathrm{max}}=0.5\ h \mathrm{Mpc}^{-1}$.
\begin{table*}[ht!]
\centering
\begin{tabular}{ccccccc}
\hline
\multicolumn{7}{c}{Marginalized Fisher Constraints}                     \\ \hline
Probes      & $\Omega_m$ & $\Omega_b$ & $h$   & $n_s$ & $\sigma_8$ & $\mathcolorbox{white}{M_\nu \ [\mathrm{eV}]}$ \\
$\delta_{cb}$    & 0.063      & 0.018      & 0.19  & 0.13  & 0.010 & 1.7     \\
$\delta_{r}$     & 0.035      & 0.0097     & 0.11  & 0.072 & 0.059 & 0.99 \\
$\delta_{nr}$    & 0.023      & 0.0066     & 0.073 & 0.049 & 0.040 & 0.67    \\
$\delta_L$       & 0.017     & 0.0055      & 0.054 & 0.027 & 0.034 & 0.59    \\
Improvement & 2.72       & 2.65       & 2.64  & 2.65  & 2.49  & 2.54    \\ \hline
\end{tabular}
\caption{ Marginalized $1\sigma$ errors of cosmological parameters for $k_{\mathrm{max}} = 0.5 \ h\mathrm{Mpc}^{-1}$.We list the constraints obtained from the power spectrum of $\delta_{cb}$, $\delta_{r}$, $\delta_{nr}$, and $\delta_L$, and the multiplicative improvement in the constraints with nonlinear reconstruction, compared to those from the power spectrum of $\delta_{cb}$ field alone.
The signal to noise ratio of neutrino mass measurement, $\propto1/\sigma(M_\nu)$, is only $12\%$ lower than the linear power spectrum, i.e., the theoretical upper limit achievable with the $\delta_{cb}$ field.
\label{tab:Constraint}}
\end{table*}
We find that nonlinear reconstruction can achieve a significant level of improvement on all 6 parameters and the improvements to be a factor of 2.72, 2.65, 2.64, 2.65, 2.49, 2.54 for $\Omega_m, \Omega_b, h, n_s, \sigma_8$ and $M_\nu$, respectively.
Thus, the constraining power of the reconstructed power spectrum $P_{\delta_{nr}}$, i.e., the signal to noise ratio of neutrino mass measurement $\propto1/\sigma(M_\nu)$, is only $12\%$ lower than the theoretical upper limit achievable with the $\delta_{cb}$ field.

We notice that the exact values of $\sigma(M_\nu)$ are slightly different between our results and \citet[][]{2022PhRvD.105l3510B}.
We have confirmed that this is due to the different fiducial cosmology adopted to compute the derivatives to neutrinos.
Our fiducial cosmology is for a Universe with massless neutrinos, while \citet[][]{2022PhRvD.105l3510B} have used a fiducial neutrino mass $M_\nu=0.05\ \mr{eV}$ when computing the linear theory constraints.
Nevertheless, we find that the relative improvements are not sensitive to these numerical details and thus this does not impact our major conclusions presented above, i.e., nonlinear reconstruction can recover almost all the neutrino information, nearly reaching the theoretical upper limit.

\subsection{Test with the theoretical model}

\begin{table}[ht!]
\centering
\begin{tabular}{ccccccc}
\hline
\multicolumn{7}{c}{Improvement}                     \\ \hline
Probes      & $\Omega_m$ & $\Omega_b$ & $h$   & $n_s$ & $\sigma_8$ & $M_\nu $ \\
$\delta_{nr}$   & 2.72       & 2.65       & 2.64  & 2.65  & 2.49  & 2.54 \\
Model     & 1.63      & 1.78     & 1.49  & 1.00 & 1.95 & 2.00 \\   \hline
\end{tabular}
\caption{Multiplicative improvement using the reconstructed power spectrum and the theoretical model for $k_{\mathrm{max}} = 0.5 \ h \mathrm{Mpc}^{-1}$.}
\label{table:Improvement}
\end{table}

In the above analysis, we have used the reconstructed power spectrum from simulations for computing the covariance matrix and derivatives to cosmological parameters.
This requires an accurate modeling of the reconstructed power spectrum, which might be achieved with the emulator constructed from simulations \citep[see e.g.,][for a recent exploration]{2023arXiv231105848W}. 

When using the perturbation theory approach, usually we need to use more parameters to describe the nonlinear effects in the reconstructed power spectrum, such as the simple model we introduced earlier.
The constraints on cosmological parameters will degrade after marginalizing over these nuisance parameters due to possible degeneracy between parameters.
We can employ the simplified model in Eq.~(\ref{eq:pk_theory}) to calculate the Fisher information. 
In this approach, we utilize the reconstructed power spectrum covariance matrix in the simulations, while using Eq.~(\ref{eq:pk_theory}) for derivative computations.
In this way, we include the damping scale $\Sigma$ and noise power spectrum $P_{\mr{N}}$ in the Fisher analysis.

In Table~\ref{table:Improvement}, we list the multiplicative improvement using the reconstructed power spectrum from simulations and the theoretical model for
$k_{\mathrm{max}}=0.5\ h \mathrm{Mpc}^{-1}$. 
We see that the improvement of the constraint on neutrino mass is $\sim20\%$ lower than using the power spectrum from simulation.
However, nonlinear reconstruction can still achieve a substantial improvement, by a factor of 2, compared to the original nonlinear power spectrum.
The constraints on other parameters are not as good as the neutrino mass, which is due to the degeneracy between the cosmological and nuisance parameters, especially the spectrum index $n_s$, since both the spectrum index and the damping scale change the power spectrum shape.

\section{Discussion and Conclusion}
\label{sec:Discussion}

In this paper, we apply nonlinear reconstruction to the CDM density fields\ from simulations and find the parameter constraints are much tighter than those from the power spectrum of $\delta_{cb}$ directly, with the improvement to be a factor of $\sim2$ for each parameter.
A salient conclusion is that the constraint on neutrino mass can nearly reach that of the linear power spectrum, with a signal-to-noise ratio only $12\%$ lower than the theoretical best result, making reconstruction an efficient way to extract cosmological information from the nonlinear galaxy survey data.

We have assumed baryons trace CDM when using the CDM field as an approximation of the CDM+baryon distribution.
This is a valid approximation on large and intermediate scales, $k\lesssim0.5\ h\mr{Mpc}^{-1}$, where the baryonic effects are subdominant \citep[see e.g.][]{2019OJAp....2E...4C,2021ApJ...915...71V}.
The analysis with $\delta_{cb}$ corresponds to an unbiased tracer with negligible shot noise, $\delta_g=b\delta_{cb}$ with $b=1$.
Therefore, our constraints on $M_\nu$ should be interpreted as maximum information content in principle.
Higher order biasing contributes tens of percent of the total galaxy power spectrum at $k\sim0.5\ h\mr{Mpc}^{-1}$ \citep[see e.g.,][]{2019PhRvD.100d3514S} and the importance of shot noise depends on the specific surveys. 
In future work, we plan to extend the analysis to simulated galaxy samples, which are more representative of real data.


In redshift space, more information regarding neutrinos is available, since the redshift space distortions is induced by the peculiar velocity, which is sourced by the total matter field and thus is sensitive to the neutrino overdensity $\delta_\nu$ \citep[][]{2022PhRvD.105l3510B}.
Therefore, it is possible to obtain better constraints in redshift space, which we plan to explore in future.

To obtain an unbiased inference of cosmological parameters, it is necessary to have an accurate power spectrum model for reconstruction.
The modeling of standard reconstruction has been explored using both the Lagrangian perturbation theory \citep[][]{2009PhRvD..80l3501N,2009PhRvD..79f3523P,2015MNRAS.450.3822W,2016MNRAS.460.2453S,2019JCAP...09..017C} and standard perturbation theory \citep[][]{2017PhRvD..96d3513H,2020PhRvD.101d3510H}.
Recently, \citet[][]{,2021PhRvD.104l3508O, 2022arXiv221107960O} have explored the modeling of nonlinear displacement using Lagrangian perturbation theory.
The simulation-based method such as power spectrum emulator might be needed on nonlinear scales, where the perturbation theory methods are no longer available.

We have exclusively considered the nonlinear reconstruction methods based on solving the full nonlinear displacement field \citep[][]{2017PhRvD..96l3502Z,2017PhRvD..96b3505S,2018PhRvD..97b3505S}.
The applications to halo fields and redshift-space distortions have been studied extensively \citep[see e.g.,][]{2017ApJ...847..110Y,2018PhRvD..97d3502Z,2019ApJ...870..116W,2019MNRAS.483.5267B,2020MNRAS.497.3451W,2021ApJS..254....4L,2022MNRAS.511.1557S}.
There are many other reconstruction methods based on different principles, e.g., iterative methods \citep[][]{2018MNRAS.478.1866H,2019MNRAS.482.5685H}, optimization techniques \citep[][]{2017JCAP...12..009S,2018JCAP...07..043F,2018JCAP...10..028M},  forwarding modeling \citep[][]{2013MNRAS.432..894J,2014ApJ...794...94W}, effective field theory based likelihood \citep[][]{2019JCAP...01..042S,2020JCAP...01..029E}, fast action minimization \citep[][]{2000MNRAS.313..587N,2019MNRAS.484.3818S}, machine learning \citep[][]{2021MNRAS.501.1499M,2022arXiv220712511S,2023arXiv230507018F}, etc, which could also provide an improvement in the cosmological parameter constraints.

While for the current galaxy samples such as BOSS \citep[][]{2017MNRAS.470.2617A} and eBOSS \citep[][]{2021PhRvD.103h3533A}, the nonlinear reconstruction methods have a similar efficiency as standard reconstruction because of the higher shot noise. 
The near term and upcoming surveys such as DESI \citep[][]{2016arXiv161100036D}, Euclid \citep[][]{2018LRR....21....2A,2020A&A...642A.191E}, MegaMapper \citep[][]{2022arXiv220904322S}, etc, will generally have a much higher galaxy number density and therefore nonlinear reconstruction can achieve a better performance with the lower shot noise.
We expect that reconstruction will be a promising method to extract cosmological information from future galaxy surveys.

\begin{acknowledgments}
We acknowledge Adrian Bayer and Gong-Bo Zhao for enlightening discussions about the Fisher matrix calculation.
Hong-Ming Zhu receives support from the National Natural Science Foundation of China (NSFC) [Grant No.~11890691] and the Natural Sciences and Engineering Research Council of Canada (NSERC) [funding reference number CITA 490888-16].
Computations were performed on the Niagara supercomputer at the SciNet HPC Consortium and the SOSCIP Consortium's CPU computing platform \citep[][]{10.1145/3332186.3332195,Loken_2010}.
SciNet is funded by: the Canada Foundation for Innovation; the Government of Ontario; Ontario Research Fund - Research Excellence; and the University of Toronto.
SOSCIP is funded by the Federal Economic Development Agency of Southern Ontario, the Province of Ontario, IBM Canada Ltd., Ontario Centres of Excellence, Mitacs and 15 Ontario academic member institutions.
\end{acknowledgments}

\vspace{5mm}

\appendix

\section{Convergence Test}
\label{sec:Convergence}

\begin{figure}[ht!]
    \centering
    \includegraphics[width=\columnwidth]{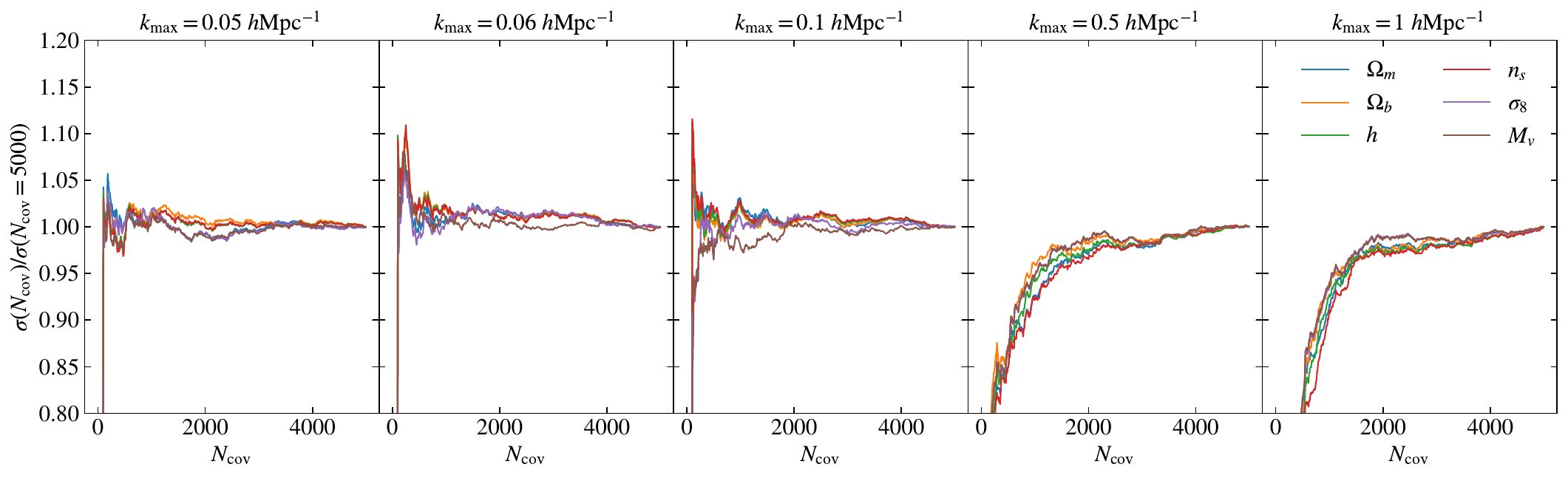}
    \caption{Marginalized $1\sigma$ error as a function of the number of simulations used to calculate the covariance matrix for nonlinear reconstruction. 
    From left to right, we show the results for different $k_{\mathrm{max}}$.
    For $N_{\mr{cov}}=5000$, a tight convergence is achieved.
    }
    \label{fig:Convergence_Cov_DisIter}
\end{figure}

In this appendix, we present the convergence tests of our results.
In Figure~\ref{fig:Convergence_Cov_DisIter}, 
we show the marginalized $1\sigma$ errors on cosmological parameters as a function of the number of simulations used to estimate the covariance matrix for nonlinear reconstruction.
The results are presented as the ratio of the errors computed using $N_{\mr{cov}}$ simulations and $5000$ simulations.
The derivatives are computed using 500 pairs of simulations.
From left to right, we show the convergence of marginalized constraints for different $k_{\mathrm{max}}$.
We see that the results converge well for all scales from $0.05\ h\mr{Mpc}^{-1}$ to $1\ h\mr{Mpc}^{-1}$, with a small error at percent level.

\begin{figure*}[ht!]
    \centering
    \includegraphics[width=\textwidth]{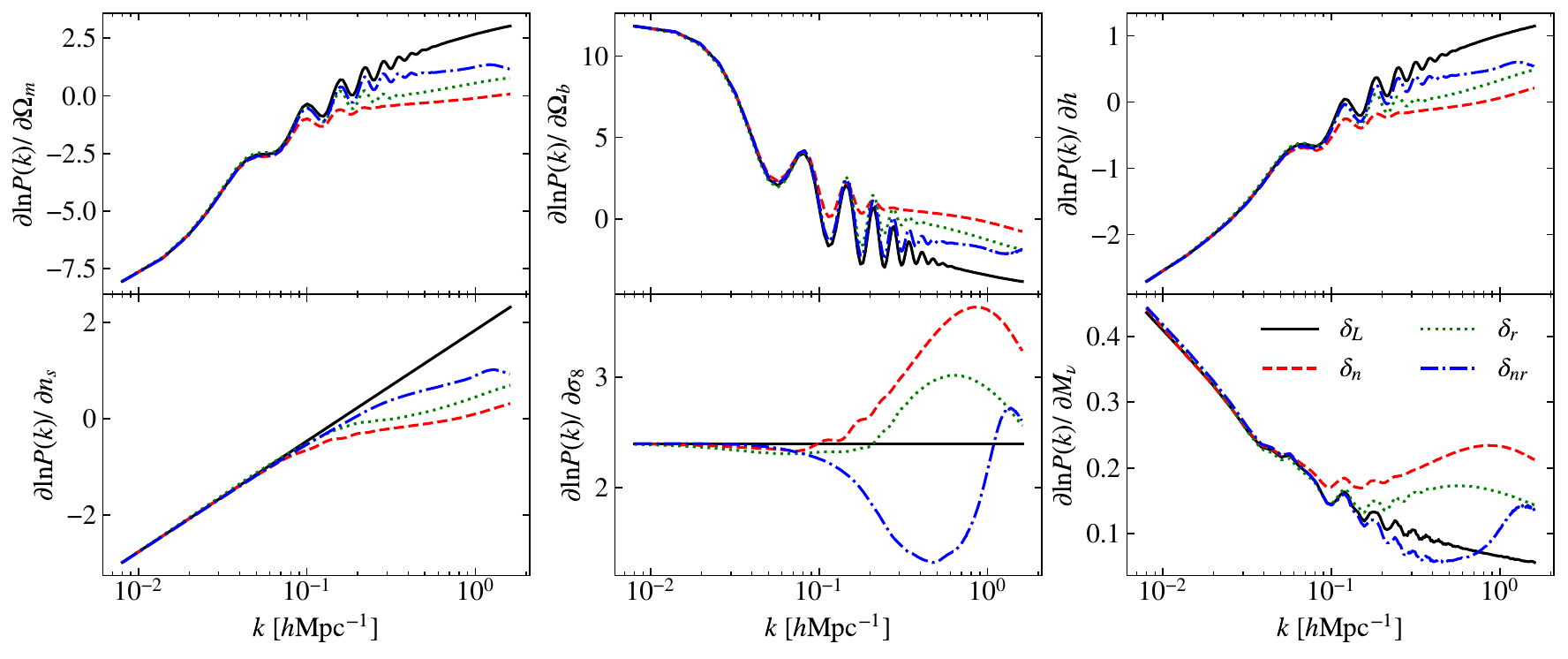}
    \caption{The derivatives of the power spectrum to cosmological parameters. 
    The derivatives are well converged using 500 pairs of simulations.
\label{fig:Deri}}
\end{figure*}

In Fig.~\ref{fig:Deri},
we show the derivatives of power spectrum with respect to cosmological parameters.
The derivatives for the power spectrum show good convergence for 500 pairs of simulations.

\begin{figure}[ht!]
    \centering
    \includegraphics[width=\columnwidth]{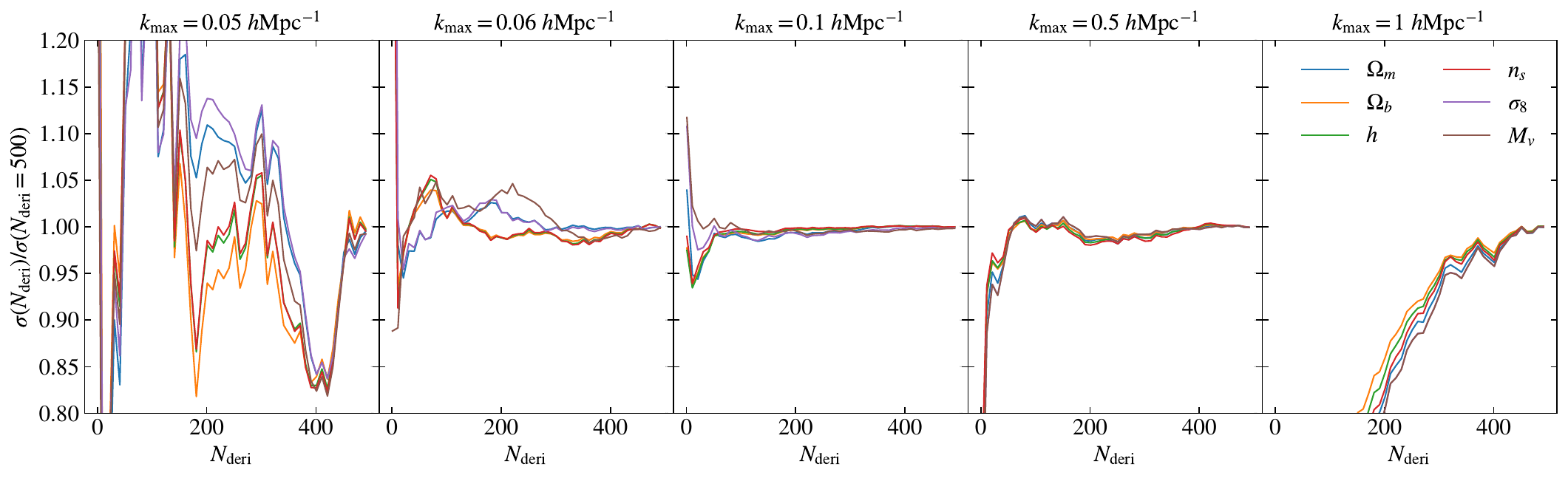}
    \caption{Marginalized $1\sigma$ error as a function of the number of simulation pairs used to calculate the partial derivatives for nonlinear reconstruction. 
    From left to right, we show the results for different $k_{\mathrm{max}}$. For $N_{\mr{deri}}=500$, a good convergence is achieved for $k_{\mathrm{max}}$ from $0.06 \ h \mathrm{Mpc}^{-1}$ to $0.5 \ h \mathrm{Mpc}^{-1}$.
\label{fig:Convergence_Deri_DisIter}}
\end{figure}

Figure~\ref{fig:Convergence_Deri_DisIter}
shows the marginalized $1\sigma$ errors with nonlinear reconstruction on the cosmological parameters as a function of the simulation pairs to compute the derivatives.
The results are presented as the ratio between the errors calculated with $N_{\mr{Deri}}$ pairs and 500 pairs of simulations.
The covariance matrix is computed using 5000 simulations.
For the derivatives, a good convergence is achieved for $k_{\mathrm{max}}$ from $0.06 \ h\mathrm{Mpc}^{-1}$ to $0.5 \ h\mathrm{Mpc}^{-1}$.
For $k_{\mathrm{max}} = 0.05 \ h\mathrm{Mpc}^{-1}$, the fractional error is about $20\%$ when using 500 simulation pairs.
There are not enough modes on these large scales to get converged results with the simulations available.
A higher accuracy power spectrum estimation is required to obtain reliable results at $k_{\mathrm{max}} = 0.05 \ h\mathrm{Mpc}^{-1}$, but it is fine since the dominant constraining power comes from small-scale modes.
For $k_{\mathrm{max}} = 0.06 \ h\mathrm{Mpc}^{-1}$, the results become much more stable.
The fractional error is under $2\%$ when using 500 simulation pairs to obtain partial derivatives.
Therefore, we present the Fisher constraints from $k_{\mathrm{max}} = 0.06 \ h\mathrm{Mpc}^{-1}$ to $k_{\mathrm{max}} = 0.5 \ h\mathrm{Mpc}^{-1}$ as our default results.

\section{Confidence intervals}
\label{sec:confidence}

\begin{figure*}[ht!]
    \centering
    \includegraphics[width=0.9\textwidth]{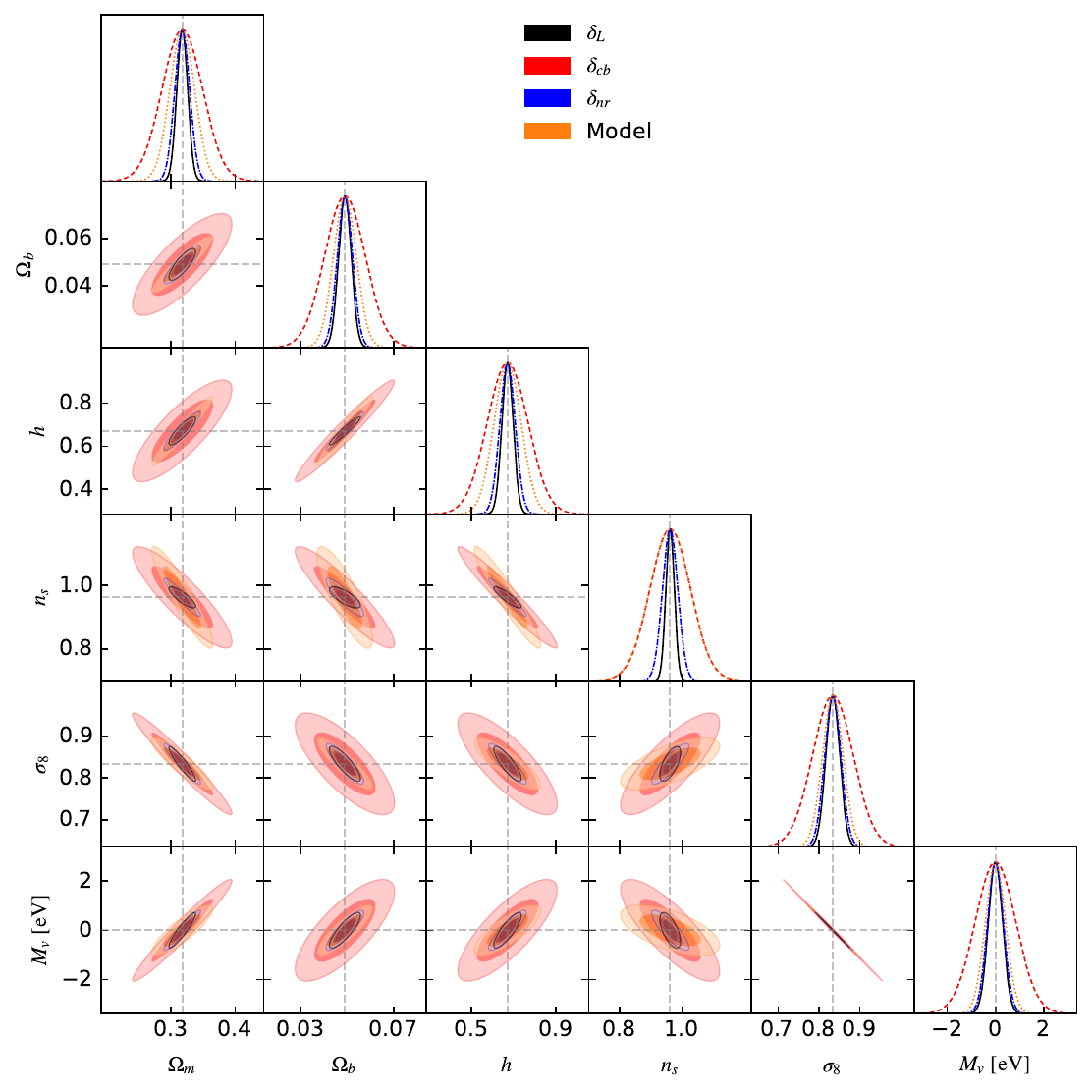}
    \caption{$68\%$ (darker shades) and $95\%$ (lighter shades) confidence contours for the cosmological parameters for $P_{\delta_L}$ (black), $P_{\delta_{cb}}$ (red), $P_{\delta_{nr}}$ (blue) and power spectrum with our theoretical model (orange) with $k_{\mathrm{max}} = 0.5 \ h \mathrm{Mpc}^{-1}$.
    The constraints are much tighter after reconstruction.
\label{fig:Contour}}
\end{figure*}

In Figure~\ref{fig:Contour},
We show the two-dimensional $68 \%$ and $95\%$ confidence intervals from the Fisher analysis for $k_{\mathrm{max}} = 0.5 \ h\mathrm{Mpc}^{-1}$.
The constraints are generally much tighter after reconstruction.
However, it does not break the strong degeneracy between $\sigma_8$ and $M_\nu$ using the reconstructed power spectrum alone.
By introducing other nonlinear statistics, a much tighter constraint can be achieved \cite[][]{2021ApJ...919...24B}.

\bibliography{sample631}{}

\begin{thebibliography}{}
\expandafter\ifx\csname natexlab\endcsname\relax\def\natexlab#1{#1}\fi
\providecommand{\url}[1]{\href{#1}{#1}}
\providecommand{\dodoi}[1]{doi:~\href{http://doi.org/#1}{\nolinkurl{#1}}}
\providecommand{\doeprint}[1]{\href{http://ascl.net/#1}{\nolinkurl{http://ascl.net/#1}}}
\providecommand{\doarXiv}[1]{\href{https://arxiv.org/abs/#1}{\nolinkurl{https://arxiv.org/abs/#1}}}

\bibitem[{{Akeson} {et~al.}(2019){Akeson}, {Armus}, {Bachelet}, {Bailey},
  {Bartusek}, {Bellini}, {Benford}, {Bennett}, {Bhattacharya}, {Bohlin},
  {Boyer}, {Bozza}, {Bryden}, {Calchi Novati}, {Carpenter}, {Casertano},
  {Choi}, {Content}, {Dayal}, {Dressler}, {Dor{\'e}}, {Fall}, {Fan}, {Fang},
  {Filippenko}, {Finkelstein}, {Foley}, {Furlanetto}, {Kalirai}, {Gaudi},
  {Gilbert}, {Girard}, {Grady}, {Greene}, {Guhathakurta}, {Heinrich},
  {Hemmati}, {Hendel}, {Henderson}, {Henning}, {Hirata}, {Ho}, {Huff},
  {Hutter}, {Jansen}, {Jha}, {Johnson}, {Jones}, {Kasdin}, {Kelly}, {Kirshner},
  {Koekemoer}, {Kruk}, {Lewis}, {Macintosh}, {Madau}, {Malhotra}, {Mandel},
  {Massara}, {Masters}, {McEnery}, {McQuinn}, {Melchior}, {Melton},
  {Mennesson}, {Peeples}, {Penny}, {Perlmutter}, {Pisani}, {Plazas}, {Poleski},
  {Postman}, {Ranc}, {Rauscher}, {Rest}, {Roberge}, {Robertson}, {Rodney},
  {Rhoads}, {Rhodes}, {Ryan}, {Sahu}, {Sand}, {Scolnic}, {Seth}, {Shvartzvald},
  {Siellez}, {Smith}, {Spergel}, {Stassun}, {Street}, {Strolger}, {Szalay},
  {Trauger}, {Troxel}, {Turnbull}, {van der Marel}, {von der Linden}, {Wang},
  {Weinberg}, {Williams}, {Windhorst}, {Wollack}, {Wu}, {Yee}, \&
  {Zimmerman}}]{2019arXiv190205569A}
{Akeson}, R., {Armus}, L., {Bachelet}, E., {et~al.} 2019, arXiv e-prints,
  arXiv:1902.05569.
\newblock \doarXiv{1902.05569}

\bibitem[{{Alam} {et~al.}(2017){Alam}, {Ata}, {Bailey}, {Beutler}, {Bizyaev},
  {Blazek}, {Bolton}, {Brownstein}, {Burden}, {Chuang}, {Comparat}, {Cuesta},
  {Dawson}, {Eisenstein}, {Escoffier}, {Gil-Mar{\'\i}n}, {Grieb}, {Hand}, {Ho},
  {Kinemuchi}, {Kirkby}, {Kitaura}, {Malanushenko}, {Malanushenko}, {Maraston},
  {McBride}, {Nichol}, {Olmstead}, {Oravetz}, {Padmanabhan},
  {Palanque-Delabrouille}, {Pan}, {Pellejero-Ibanez}, {Percival}, {Petitjean},
  {Prada}, {Price-Whelan}, {Reid}, {Rodr{\'\i}guez-Torres}, {Roe}, {Ross},
  {Ross}, {Rossi}, {Rubi{\~n}o-Mart{\'\i}n}, {Saito}, {Salazar-Albornoz},
  {Samushia}, {S{\'a}nchez}, {Satpathy}, {Schlegel}, {Schneider},
  {Sc{\'o}ccola}, {Seo}, {Sheldon}, {Simmons}, {Slosar}, {Strauss}, {Swanson},
  {Thomas}, {Tinker}, {Tojeiro}, {Maga{\~n}a}, {Vazquez}, {Verde}, {Wake},
  {Wang}, {Weinberg}, {White}, {Wood-Vasey}, {Y{\`e}che}, {Zehavi}, {Zhai}, \&
  {Zhao}}]{2017MNRAS.470.2617A}
{Alam}, S., {Ata}, M., {Bailey}, S., {et~al.} 2017, \mnras, 470, 2617,
  \dodoi{10.1093/mnras/stx721}

\bibitem[{{Alam} {et~al.}(2021){Alam}, {Aubert}, {Avila}, {Balland},
  {Bautista}, {Bershady}, {Bizyaev}, {Blanton}, {Bolton}, {Bovy}, {Brinkmann},
  {Brownstein}, {Burtin}, {Chabanier}, {Chapman}, {Choi}, {Chuang}, {Comparat},
  {Cousinou}, {Cuceu}, {Dawson}, {de la Torre}, {de Mattia}, {Agathe}, {des
  Bourboux}, {Escoffier}, {Etourneau}, {Farr}, {Font-Ribera}, {Frinchaboy},
  {Fromenteau}, {Gil-Mar{\'\i}n}, {Le Goff}, {Gonzalez-Morales},
  {Gonzalez-Perez}, {Grabowski}, {Guy}, {Hawken}, {Hou}, {Kong}, {Parker},
  {Klaene}, {Kneib}, {Lin}, {Long}, {Lyke}, {de la Macorra}, {Martini},
  {Masters}, {Mohammad}, {Moon}, {Mueller}, {Mu{\~n}oz-Guti{\'e}rrez}, {Myers},
  {Nadathur}, {Neveux}, {Newman}, {Noterdaeme}, {Oravetz}, {Oravetz},
  {Palanque-Delabrouille}, {Pan}, {Paviot}, {Percival}, {P{\'e}rez-R{\`a}fols},
  {Petitjean}, {Pieri}, {Prakash}, {Raichoor}, {Ravoux}, {Rezaie}, {Rich},
  {Ross}, {Rossi}, {Ruggeri}, {Ruhlmann-Kleider}, {S{\'a}nchez}, {S{\'a}nchez},
  {S{\'a}nchez-Gallego}, {Sayres}, {Schneider}, {Seo}, {Shafieloo}, {Slosar},
  {Smith}, {Stermer}, {Tamone}, {Tinker}, {Tojeiro}, {Vargas-Maga{\~n}a},
  {Variu}, {Wang}, {Weaver}, {Weijmans}, {Y{\`e}che}, {Zarrouk}, {Zhao},
  {Zhao}, \& {Zheng}}]{2021PhRvD.103h3533A}
{Alam}, S., {Aubert}, M., {Avila}, S., {et~al.} 2021, \prd, 103, 083533,
  \dodoi{10.1103/PhysRevD.103.083533}

\bibitem[{{Amendola} {et~al.}(2018){Amendola}, {Appleby}, {Avgoustidis},
  {Bacon}, {Baker}, {Baldi}, {Bartolo}, {Blanchard}, {Bonvin}, {Borgani},
  {Branchini}, {Burrage}, {Camera}, {Carbone}, {Casarini}, {Cropper}, {de
  Rham}, {Dietrich}, {Di Porto}, {Durrer}, {Ealet}, {Ferreira}, {Finelli},
  {Garc{\'\i}a-Bellido}, {Giannantonio}, {Guzzo}, {Heavens}, {Heisenberg},
  {Heymans}, {Hoekstra}, {Hollenstein}, {Holmes}, {Hwang}, {Jahnke},
  {Kitching}, {Koivisto}, {Kunz}, {La Vacca}, {Linder}, {March}, {Marra},
  {Martins}, {Majerotto}, {Markovic}, {Marsh}, {Marulli}, {Massey}, {Mellier},
  {Montanari}, {Mota}, {Nunes}, {Percival}, {Pettorino}, {Porciani},
  {Quercellini}, {Read}, {Rinaldi}, {Sapone}, {Sawicki}, {Scaramella},
  {Skordis}, {Simpson}, {Taylor}, {Thomas}, {Trotta}, {Verde}, {Vernizzi},
  {Vollmer}, {Wang}, {Weller}, \& {Zlosnik}}]{2018LRR....21....2A}
{Amendola}, L., {Appleby}, S., {Avgoustidis}, A., {et~al.} 2018, Living Reviews
  in Relativity, 21, 2, \dodoi{10.1007/s41114-017-0010-3}

\bibitem[{{Bayer} {et~al.}(2022){Bayer}, {Banerjee}, \&
  {Seljak}}]{2022PhRvD.105l3510B}
{Bayer}, A.~E., {Banerjee}, A., \& {Seljak}, U. 2022, \prd, 105, 123510,
  \dodoi{10.1103/PhysRevD.105.123510}

\bibitem[{{Bayer} {et~al.}(2021){Bayer}, {Villaescusa-Navarro}, {Massara},
  {Liu}, {Spergel}, {Verde}, {Wandelt}, {Viel}, \& {Ho}}]{2021ApJ...919...24B}
{Bayer}, A.~E., {Villaescusa-Navarro}, F., {Massara}, E., {et~al.} 2021, \apj,
  919, 24, \dodoi{10.3847/1538-4357/ac0e91}

\bibitem[{{Birkin} {et~al.}(2019){Birkin}, {Li}, {Cautun}, \&
  {Shi}}]{2019MNRAS.483.5267B}
{Birkin}, J., {Li}, B., {Cautun}, M., \& {Shi}, Y. 2019, \mnras, 483, 5267,
  \dodoi{10.1093/mnras/sty3365}

\bibitem[{Bond {et~al.}(1980)Bond, Efstathiou, \&
  Silk}]{1980PhysRevLett.45.1980}
Bond, J.~R., Efstathiou, G., \& Silk, J. 1980, \prl, 45, 1980,
  \dodoi{10.1103/PhysRevLett.45.1980}

\bibitem[{{Carron}(2013)}]{2013A&A...551A..88C}
{Carron}, J. 2013, \aap, 551, A88, \dodoi{10.1051/0004-6361/201220538}

\bibitem[{{Castorina} {et~al.}(2015){Castorina}, {Carbone}, {Bel}, {Sefusatti},
  \& {Dolag}}]{2015JCAP...07..043C}
{Castorina}, E., {Carbone}, C., {Bel}, J., {Sefusatti}, E., \& {Dolag}, K.
  2015, \jcap, 2015, 043, \dodoi{10.1088/1475-7516/2015/07/043}

\bibitem[{{Chen} {et~al.}(2019){Chen}, {Vlah}, \&
  {White}}]{2019JCAP...09..017C}
{Chen}, S.-F., {Vlah}, Z., \& {White}, M. 2019, \jcap, 2019, 017,
  \dodoi{10.1088/1475-7516/2019/09/017}

\bibitem[{{Chisari} {et~al.}(2019){Chisari}, {Mead}, {Joudaki}, {Ferreira},
  {Schneider}, {Mohr}, {Tr{\"o}ster}, {Alonso}, {McCarthy}, {Martin-Alvarez},
  {Devriendt}, {Slyz}, \& {van Daalen}}]{2019OJAp....2E...4C}
{Chisari}, N.~E., {Mead}, A.~J., {Joudaki}, S., {et~al.} 2019, The Open Journal
  of Astrophysics, 2, 4, \dodoi{10.21105/astro.1905.06082}

\bibitem[{{DESI Collaboration} {et~al.}(2016){DESI Collaboration}, {Aghamousa},
  {Aguilar}, {Ahlen}, {Alam}, {Allen}, {Allende Prieto}, {Annis}, {Bailey},
  {Balland}, {Ballester}, {Baltay}, {Beaufore}, {Bebek}, {Beers}, {Bell},
  {Bernal}, {Besuner}, {Beutler}, {Blake}, {Bleuler}, {Blomqvist}, {Blum},
  {Bolton}, {Briceno}, {Brooks}, {Brownstein}, {Buckley-Geer}, {Burden},
  {Burtin}, {Busca}, {Cahn}, {Cai}, {Cardiel-Sas}, {Carlberg}, {Carton},
  {Casas}, {Castander}, {Cervantes-Cota}, {Claybaugh}, {Close}, {Coker},
  {Cole}, {Comparat}, {Cooper}, {Cousinou}, {Crocce}, {Cuby}, {Cunningham},
  {Davis}, {Dawson}, {de la Macorra}, {De Vicente}, {Delubac}, {Derwent},
  {Dey}, {Dhungana}, {Ding}, {Doel}, {Duan}, {Ealet}, {Edelstein},
  {Eftekharzadeh}, {Eisenstein}, {Elliott}, {Escoffier}, {Evatt}, {Fagrelius},
  {Fan}, {Fanning}, {Farahi}, {Farihi}, {Favole}, {Feng}, {Fernandez},
  {Findlay}, {Finkbeiner}, {Fitzpatrick}, {Flaugher}, {Flender}, {Font-Ribera},
  {Forero-Romero}, {Fosalba}, {Frenk}, {Fumagalli}, {Gaensicke}, {Gallo},
  {Garcia-Bellido}, {Gaztanaga}, {Pietro Gentile Fusillo}, {Gerard},
  {Gershkovich}, {Giannantonio}, {Gillet}, {Gonzalez-de-Rivera},
  {Gonzalez-Perez}, {Gott}, {Graur}, {Gutierrez}, {Guy}, {Habib}, {Heetderks},
  {Heetderks}, {Heitmann}, {Hellwing}, {Herrera}, {Ho}, {Holland}, {Honscheid},
  {Huff}, {Hutchinson}, {Huterer}, {Hwang}, {Illa Laguna}, {Ishikawa},
  {Jacobs}, {Jeffrey}, {Jelinsky}, {Jennings}, {Jiang}, {Jimenez}, {Johnson},
  {Joyce}, {Jullo}, {Juneau}, {Kama}, {Karcher}, {Karkar}, {Kehoe}, {Kennamer},
  {Kent}, {Kilbinger}, {Kim}, {Kirkby}, {Kisner}, {Kitanidis}, {Kneib},
  {Koposov}, {Kovacs}, {Koyama}, {Kremin}, {Kron}, {Kronig}, {Kueter-Young},
  {Lacey}, {Lafever}, {Lahav}, {Lambert}, {Lampton}, {Landriau}, {Lang},
  {Lauer}, {Le Goff}, {Le Guillou}, {Le Van Suu}, {Lee}, {Lee}, {Leitner},
  {Lesser}, {Levi}, {L'Huillier}, {Li}, {Liang}, {Lin}, {Linder}, {Loebman},
  {Luki{\'c}}, {Ma}, {MacCrann}, {Magneville}, {Makarem}, {Manera}, {Manser},
  {Marshall}, {Martini}, {Massey}, {Matheson}, {McCauley}, {McDonald},
  {McGreer}, {Meisner}, {Metcalfe}, {Miller}, {Miquel}, {Moustakas}, {Myers},
  {Naik}, {Newman}, {Nichol}, {Nicola}, {Nicolati da Costa}, {Nie}, {Niz},
  {Norberg}, {Nord}, {Norman}, {Nugent}, {O'Brien}, {Oh}, {Olsen}, {Padilla},
  {Padmanabhan}, {Padmanabhan}, {Palanque-Delabrouille}, {Palmese},
  {Pappalardo}, {P{\^a}ris}, {Park}, {Patej}, {Peacock}, {Peiris}, {Peng},
  {Percival}, {Perruchot}, {Pieri}, {Pogge}, {Pollack}, {Poppett}, {Prada},
  {Prakash}, {Probst}, {Rabinowitz}, {Raichoor}, {Ree}, {Refregier}, {Regal},
  {Reid}, {Reil}, {Rezaie}, {Rockosi}, {Roe}, {Ronayette}, {Roodman}, {Ross},
  {Ross}, {Rossi}, {Rozo}, {Ruhlmann-Kleider}, {Rykoff}, {Sabiu}, {Samushia},
  {Sanchez}, {Sanchez}, {Schlegel}, {Schneider}, {Schubnell}, {Secroun},
  {Seljak}, {Seo}, {Serrano}, {Shafieloo}, {Shan}, {Sharples}, {Sholl},
  {Shourt}, {Silber}, {Silva}, {Sirk}, {Slosar}, {Smith}, {Smoot}, {Som},
  {Song}, {Sprayberry}, {Staten}, {Stefanik}, {Tarle}, {Sien Tie}, {Tinker},
  {Tojeiro}, {Valdes}, {Valenzuela}, {Valluri}, {Vargas-Magana}, {Verde},
  {Walker}, {Wang}, {Wang}, {Weaver}, {Weaverdyck}, {Wechsler}, {Weinberg},
  {White}, {Yang}, {Yeche}, {Zhang}, {Zhao}, {Zheng}, {Zhou}, {Zhou}, {Zhu},
  {Zou}, \& {Zu}}]{2016arXiv161100036D}
{DESI Collaboration}, {Aghamousa}, A., {Aguilar}, J., {et~al.} 2016, arXiv
  e-prints, arXiv:1611.00036.
\newblock \doarXiv{1611.00036}

\bibitem[{{Dor{\'e}} {et~al.}(2014){Dor{\'e}}, {Bock}, {Ashby}, {Capak},
  {Cooray}, {de Putter}, {Eifler}, {Flagey}, {Gong}, {Habib}, {Heitmann},
  {Hirata}, {Jeong}, {Katti}, {Korngut}, {Krause}, {Lee}, {Masters},
  {Mauskopf}, {Melnick}, {Mennesson}, {Nguyen}, {{\"O}berg}, {Pullen},
  {Raccanelli}, {Smith}, {Song}, {Tolls}, {Unwin}, {Venumadhav}, {Viero},
  {Werner}, \& {Zemcov}}]{2014arXiv1412.4872D}
{Dor{\'e}}, O., {Bock}, J., {Ashby}, M., {et~al.} 2014, arXiv e-prints,
  arXiv:1412.4872.
\newblock \doarXiv{1412.4872}

\bibitem[{{Eisenstein} {et~al.}(2007){Eisenstein}, {Seo}, {Sirko}, \&
  {Spergel}}]{2007ApJ...664..675E}
{Eisenstein}, D.~J., {Seo}, H.-J., {Sirko}, E., \& {Spergel}, D.~N. 2007, \apj,
  664, 675, \dodoi{10.1086/518712}

\bibitem[{{Elsner} {et~al.}(2020){Elsner}, {Schmidt}, {Jasche}, {Lavaux}, \&
  {Nguyen}}]{2020JCAP...01..029E}
{Elsner}, F., {Schmidt}, F., {Jasche}, J., {Lavaux}, G., \& {Nguyen}, N.-M.
  2020, \jcap, 2020, 029, \dodoi{10.1088/1475-7516/2020/01/029}

\bibitem[{{Euclid Collaboration} {et~al.}(2020){Euclid Collaboration},
  {Blanchard}, {Camera}, {Carbone}, {Cardone}, {Casas}, {Clesse}, {Ili{\'c}},
  {Kilbinger}, {Kitching}, {Kunz}, {Lacasa}, {Linder}, {Majerotto},
  {Markovi{\v{c}}}, {Martinelli}, {Pettorino}, {Pourtsidou}, {Sakr},
  {S{\'a}nchez}, {Sapone}, {Tutusaus}, {Yahia-Cherif}, {Yankelevich},
  {Andreon}, {Aussel}, {Balaguera-Antol{\'\i}nez}, {Baldi}, {Bardelli},
  {Bender}, {Biviano}, {Bonino}, {Boucaud}, {Bozzo}, {Branchini}, {Brau-Nogue},
  {Brescia}, {Brinchmann}, {Burigana}, {Cabanac}, {Capobianco}, {Cappi},
  {Carretero}, {Carvalho}, {Casas}, {Castander}, {Castellano}, {Cavuoti},
  {Cimatti}, {Cledassou}, {Colodro-Conde}, {Congedo}, {Conselice}, {Conversi},
  {Copin}, {Corcione}, {Coupon}, {Courtois}, {Cropper}, {Da Silva}, {de la
  Torre}, {Di Ferdinando}, {Dubath}, {Ducret}, {Duncan}, {Dupac}, {Dusini},
  {Fabbian}, {Fabricius}, {Farrens}, {Fosalba}, {Fotopoulou}, {Fourmanoit},
  {Frailis}, {Franceschi}, {Franzetti}, {Fumana}, {Galeotta}, {Gillard},
  {Gillis}, {Giocoli}, {G{\'o}mez-Alvarez}, {Graci{\'a}-Carpio}, {Grupp},
  {Guzzo}, {Hoekstra}, {Hormuth}, {Israel}, {Jahnke}, {Keihanen}, {Kermiche},
  {Kirkpatrick}, {Kohley}, {Kubik}, {Kurki-Suonio}, {Ligori}, {Lilje}, {Lloro},
  {Maino}, {Maiorano}, {Marggraf}, {Martinet}, {Marulli}, {Massey},
  {Medinaceli}, {Mei}, {Mellier}, {Metcalf}, {Metge}, {Meylan}, {Moresco},
  {Moscardini}, {Munari}, {Nichol}, {Niemi}, {Nucita}, {Padilla}, {Paltani},
  {Pasian}, {Percival}, {Pires}, {Polenta}, {Poncet}, {Pozzetti}, {Racca},
  {Raison}, {Renzi}, {Rhodes}, {Romelli}, {Roncarelli}, {Rossetti}, {Saglia},
  {Schneider}, {Scottez}, {Secroun}, {Sirri}, {Stanco}, {Starck}, {Sureau},
  {Tallada-Cresp{\'\i}}, {Tavagnacco}, {Taylor}, {Tenti}, {Tereno},
  {Toledo-Moreo}, {Torradeflot}, {Valenziano}, {Vassallo}, {Verdoes Kleijn},
  {Viel}, {Wang}, {Zacchei}, {Zoubian}, \& {Zucca}}]{2020A&A...642A.191E}
{Euclid Collaboration}, {Blanchard}, A., {Camera}, S., {et~al.} 2020, \aap,
  642, A191, \dodoi{10.1051/0004-6361/202038071}

\bibitem[{{Feng} {et~al.}(2018){Feng}, {Seljak}, \&
  {Zaldarriaga}}]{2018JCAP...07..043F}
{Feng}, Y., {Seljak}, U., \& {Zaldarriaga}, M. 2018, \jcap, 2018, 043,
  \dodoi{10.1088/1475-7516/2018/07/043}

\bibitem[{{Ferraro} {et~al.}(2022){Ferraro}, {Sailer}, {Slosar}, \&
  {White}}]{2022arXiv220307506F}
{Ferraro}, S., {Sailer}, N., {Slosar}, A., \& {White}, M. 2022, arXiv e-prints,
  arXiv:2203.07506.
\newblock \doarXiv{2203.07506}

\bibitem[{{Fisher}(1925)}]{1925PCPS...22..700F}
{Fisher}, R.~A. 1925, Proceedings of the Cambridge Philosophical Society, 22,
  700, \dodoi{10.1017/S0305004100009580}

\bibitem[{{Fl{\"o}ss} \& {Meerburg}(2023)}]{2023arXiv230507018F}
{Fl{\"o}ss}, T., \& {Meerburg}, P.~D. 2023, arXiv e-prints, arXiv:2305.07018,
  \dodoi{10.48550/arXiv.2305.07018}

\bibitem[{{Hada} \& {Eisenstein}(2018)}]{2018MNRAS.478.1866H}
{Hada}, R., \& {Eisenstein}, D.~J. 2018, \mnras, 478, 1866,
  \dodoi{10.1093/mnras/sty1203}

\bibitem[{{Hada} \& {Eisenstein}(2019)}]{2019MNRAS.482.5685H}
---. 2019, \mnras, 482, 5685, \dodoi{10.1093/mnras/sty3137}

\bibitem[{{Hahn} {et~al.}(2020){Hahn}, {Villaescusa-Navarro}, {Castorina}, \&
  {Scoccimarro}}]{2020JCAP...03..040H}
{Hahn}, C., {Villaescusa-Navarro}, F., {Castorina}, E., \& {Scoccimarro}, R.
  2020, \jcap, 2020, 040, \dodoi{10.1088/1475-7516/2020/03/040}

\bibitem[{{Hikage} {et~al.}(2017){Hikage}, {Koyama}, \&
  {Heavens}}]{2017PhRvD..96d3513H}
{Hikage}, C., {Koyama}, K., \& {Heavens}, A. 2017, \prd, 96, 043513,
  \dodoi{10.1103/PhysRevD.96.043513}

\bibitem[{{Hikage} {et~al.}(2020){Hikage}, {Koyama}, \&
  {Takahashi}}]{2020PhRvD.101d3510H}
{Hikage}, C., {Koyama}, K., \& {Takahashi}, R. 2020, \prd, 101, 043510,
  \dodoi{10.1103/PhysRevD.101.043510}

\bibitem[{{Jasche} \& {Wandelt}(2013)}]{2013MNRAS.432..894J}
{Jasche}, J., \& {Wandelt}, B.~D. 2013, \mnras, 432, 894,
  \dodoi{10.1093/mnras/stt449}

\bibitem[{{Lesgourgues} \& {Pastor}(2006)}]{2006PhR...429..307L}
{Lesgourgues}, J., \& {Pastor}, S. 2006, \physrep, 429, 307,
  \dodoi{10.1016/j.physrep.2006.04.001}

\bibitem[{{Lewis} {et~al.}(2000){Lewis}, {Challinor}, \&
  {Lasenby}}]{2000ApJ...538..473L}
{Lewis}, A., {Challinor}, A., \& {Lasenby}, A. 2000, \apj, 538, 473,
  \dodoi{10.1086/309179}

\bibitem[{{Liu} {et~al.}(2021){Liu}, {Yu}, \& {Li}}]{2021ApJS..254....4L}
{Liu}, Y., {Yu}, Y., \& {Li}, B. 2021, \apjs, 254, 4,
  \dodoi{10.3847/1538-4365/abe868}

\bibitem[{Loken {et~al.}(2010)Loken, Gruner, Groer, Peltier, Bunn, Craig,
  Henriques, Dempsey, Yu, Chen, Dursi, Chong, Northrup, Pinto, Knecht, \&
  Zon}]{Loken_2010}
Loken, C., Gruner, D., Groer, L., {et~al.} 2010, Journal of Physics: Conference
  Series, 256, 012026, \dodoi{10.1088/1742-6596/256/1/012026}

\bibitem[{{LSST Science Collaboration} {et~al.}(2009){LSST Science
  Collaboration}, {Abell}, {Allison}, {Anderson}, {Andrew}, {Angel}, {Armus},
  {Arnett}, {Asztalos}, {Axelrod}, {Bailey}, {Ballantyne}, {Bankert},
  {Barkhouse}, {Barr}, {Barrientos}, {Barth}, {Bartlett}, {Becker}, {Becla},
  {Beers}, {Bernstein}, {Biswas}, {Blanton}, {Bloom}, {Bochanski}, {Boeshaar},
  {Borne}, {Bradac}, {Brandt}, {Bridge}, {Brown}, {Brunner}, {Bullock},
  {Burgasser}, {Burge}, {Burke}, {Cargile}, {Chandrasekharan}, {Chartas},
  {Chesley}, {Chu}, {Cinabro}, {Claire}, {Claver}, {Clowe}, {Connolly}, {Cook},
  {Cooke}, {Cooray}, {Covey}, {Culliton}, {de Jong}, {de Vries}, {Debattista},
  {Delgado}, {Dell'Antonio}, {Dhital}, {Di Stefano}, {Dickinson}, {Dilday},
  {Djorgovski}, {Dobler}, {Donalek}, {Dubois-Felsmann}, {Durech},
  {Eliasdottir}, {Eracleous}, {Eyer}, {Falco}, {Fan}, {Fassnacht}, {Ferguson},
  {Fernandez}, {Fields}, {Finkbeiner}, {Figueroa}, {Fox}, {Francke}, {Frank},
  {Frieman}, {Fromenteau}, {Furqan}, {Galaz}, {Gal-Yam}, {Garnavich},
  {Gawiser}, {Geary}, {Gee}, {Gibson}, {Gilmore}, {Grace}, {Green}, {Gressler},
  {Grillmair}, {Habib}, {Haggerty}, {Hamuy}, {Harris}, {Hawley}, {Heavens},
  {Hebb}, {Henry}, {Hileman}, {Hilton}, {Hoadley}, {Holberg}, {Holman},
  {Howell}, {Infante}, {Ivezic}, {Jacoby}, {Jain}, {R}, {Jedicke}, {Jee},
  {Garrett Jernigan}, {Jha}, {Johnston}, {Jones}, {Juric}, {Kaasalainen},
  {Styliani}, {Kafka}, {Kahn}, {Kaib}, {Kalirai}, {Kantor}, {Kasliwal},
  {Keeton}, {Kessler}, {Knezevic}, {Kowalski}, {Krabbendam}, {Krughoff},
  {Kulkarni}, {Kuhlman}, {Lacy}, {Lepine}, {Liang}, {Lien}, {Lira}, {Long},
  {Lorenz}, {Lotz}, {Lupton}, {Lutz}, {Macri}, {Mahabal}, {Mandelbaum},
  {Marshall}, {May}, {McGehee}, {Meadows}, {Meert}, {Milani}, {Miller},
  {Miller}, {Mills}, {Minniti}, {Monet}, {Mukadam}, {Nakar}, {Neill}, {Newman},
  {Nikolaev}, {Nordby}, {O'Connor}, {Oguri}, {Oliver}, {Olivier}, {Olsen},
  {Olsen}, {Olszewski}, {Oluseyi}, {Padilla}, {Parker}, {Pepper}, {Peterson},
  {Petry}, {Pinto}, {Pizagno}, {Popescu}, {Prsa}, {Radcka}, {Raddick},
  {Rasmussen}, {Rau}, {Rho}, {Rhoads}, {Richards}, {Ridgway}, {Robertson},
  {Roskar}, {Saha}, {Sarajedini}, {Scannapieco}, {Schalk}, {Schindler},
  {Schmidt}, {Schmidt}, {Schneider}, {Schumacher}, {Scranton}, {Sebag},
  {Seppala}, {Shemmer}, {Simon}, {Sivertz}, {Smith}, {Allyn Smith}, {Smith},
  {Spitz}, {Stanford}, {Stassun}, {Strader}, {Strauss}, {Stubbs}, {Sweeney},
  {Szalay}, {Szkody}, {Takada}, {Thorman}, {Trilling}, {Trimble}, {Tyson}, {Van
  Berg}, {Vanden Berk}, {VanderPlas}, {Verde}, {Vrsnak}, {Walkowicz},
  {Wandelt}, {Wang}, {Wang}, {Warner}, {Wechsler}, {West}, {Wiecha},
  {Williams}, {Willman}, {Wittman}, {Wolff}, {Wood-Vasey}, {Wozniak}, {Young},
  {Zentner}, \& {Zhan}}]{2009arXiv0912.0201L}
{LSST Science Collaboration}, {Abell}, P.~A., {Allison}, J., {et~al.} 2009,
  arXiv e-prints, arXiv:0912.0201.
\newblock \doarXiv{0912.0201}

\bibitem[{{Mao} {et~al.}(2021){Mao}, {Wang}, {Li}, {Cai}, {Falck}, {Neyrinck},
  \& {Szalay}}]{2021MNRAS.501.1499M}
{Mao}, T.-X., {Wang}, J., {Li}, B., {et~al.} 2021, \mnras, 501, 1499,
  \dodoi{10.1093/mnras/staa3741}

\bibitem[{{Massara} {et~al.}(2021){Massara}, {Villaescusa-Navarro}, {Ho},
  {Dalal}, \& {Spergel}}]{2021PhRvL.126a1301M}
{Massara}, E., {Villaescusa-Navarro}, F., {Ho}, S., {Dalal}, N., \& {Spergel},
  D.~N. 2021, \prl, 126, 011301,
  \dodoi{10.1103/PhysRevLett.126.01130110.48550/arXiv.2001.11024}

\bibitem[{{Meiksin} \& {White}(1999)}]{1999MNRAS.308.1179M}
{Meiksin}, A., \& {White}, M. 1999, \mnras, 308, 1179,
  \dodoi{10.1046/j.1365-8711.1999.02825.x}

\bibitem[{{Modi} {et~al.}(2018){Modi}, {Feng}, \&
  {Seljak}}]{2018JCAP...10..028M}
{Modi}, C., {Feng}, Y., \& {Seljak}, U. 2018, \jcap, 2018, 028,
  \dodoi{10.1088/1475-7516/2018/10/028}

\bibitem[{{Noh} {et~al.}(2009){Noh}, {White}, \&
  {Padmanabhan}}]{2009PhRvD..80l3501N}
{Noh}, Y., {White}, M., \& {Padmanabhan}, N. 2009, \prd, 80, 123501,
  \dodoi{10.1103/PhysRevD.80.123501}

\bibitem[{{Nusser} \& {Branchini}(2000)}]{2000MNRAS.313..587N}
{Nusser}, A., \& {Branchini}, E. 2000, \mnras, 313, 587,
  \dodoi{10.1046/j.1365-8711.2000.03261.x}

\bibitem[{{Ota} {et~al.}(2021){Ota}, {Seo}, {Saito}, \&
  {Beutler}}]{2021PhRvD.104l3508O}
{Ota}, A., {Seo}, H.-J., {Saito}, S., \& {Beutler}, F. 2021, \prd, 104, 123508,
  \dodoi{10.1103/PhysRevD.104.123508}

\bibitem[{{Ota} {et~al.}(2022){Ota}, {Seo}, {Saito}, \&
  {Beutler}}]{2022arXiv221107960O}
---. 2022, arXiv e-prints, arXiv:2211.07960.
\newblock \doarXiv{2211.07960}

\bibitem[{{Padmanabhan} {et~al.}(2009){Padmanabhan}, {White}, \&
  {Cohn}}]{2009PhRvD..79f3523P}
{Padmanabhan}, N., {White}, M., \& {Cohn}, J.~D. 2009, \prd, 79, 063523,
  \dodoi{10.1103/PhysRevD.79.063523}

\bibitem[{{Padmanabhan} {et~al.}(2012){Padmanabhan}, {Xu}, {Eisenstein},
  {Scalzo}, {Cuesta}, {Mehta}, \& {Kazin}}]{2012MNRAS.427.2132P}
{Padmanabhan}, N., {Xu}, X., {Eisenstein}, D.~J., {et~al.} 2012, \mnras, 427,
  2132, \dodoi{10.1111/j.1365-2966.2012.21888.x}

\bibitem[{{Paillas} {et~al.}(2022){Paillas}, {Cuesta-Lazaro}, {Zarrouk}, {Cai},
  {Percival}, {Nadathur}, {Pinon}, {de Mattia}, \&
  {Beutler}}]{2022arXiv220904310P}
{Paillas}, E., {Cuesta-Lazaro}, C., {Zarrouk}, P., {et~al.} 2022, arXiv
  e-prints, arXiv:2209.04310.
\newblock \doarXiv{2209.04310}

\bibitem[{{Pan} {et~al.}(2017){Pan}, {Pen}, {Inman}, \&
  {Yu}}]{2017MNRAS.469.1968P}
{Pan}, Q., {Pen}, U.-L., {Inman}, D., \& {Yu}, H.-R. 2017, \mnras, 469, 1968,
  \dodoi{10.1093/mnras/stx774}

\bibitem[{{Percival} {et~al.}(2022){Percival}, {Friedrich}, {Sellentin}, \&
  {Heavens}}]{2022MNRAS.510.3207P}
{Percival}, W.~J., {Friedrich}, O., {Sellentin}, E., \& {Heavens}, A. 2022,
  \mnras, 510, 3207, \dodoi{10.1093/mnras/stab354010.48550/arXiv.2108.10402}

\bibitem[{Ponce {et~al.}(2019)Ponce, van Zon, Northrup, Gruner, Chen, Ertinaz,
  Fedoseev, Groer, Mao, Mundim, Nolta, Pinto, Saldarriaga, Slavnic, Spence, Yu,
  \& Peltier}]{10.1145/3332186.3332195}
Ponce, M., van Zon, R., Northrup, S., {et~al.} 2019, in Proceedings of the
  Practice and Experience in Advanced Research Computing on Rise of the
  Machines (Learning), PEARC '19 (New York, NY, USA: Association for Computing
  Machinery), \dodoi{10.1145/3332186.3332195}

\bibitem[{{Rimes} \& {Hamilton}(2005)}]{2005MNRAS.360L..82R}
{Rimes}, C.~D., \& {Hamilton}, A. J.~S. 2005, \mnras, 360, L82,
  \dodoi{10.1111/j.1745-3933.2005.00051.x}

\bibitem[{{Rimes} \& {Hamilton}(2006)}]{2006MNRAS.371.1205R}
---. 2006, \mnras, 371, 1205, \dodoi{10.1111/j.1365-2966.2006.10710.x}

\bibitem[{{Sailer} {et~al.}(2021){Sailer}, {Castorina}, {Ferraro}, \&
  {White}}]{2021JCAP...12..049S}
{Sailer}, N., {Castorina}, E., {Ferraro}, S., \& {White}, M. 2021, \jcap, 2021,
  049, \dodoi{10.1088/1475-7516/2021/12/049}

\bibitem[{{Sarpa} {et~al.}(2019){Sarpa}, {Schimd}, {Branchini}, \&
  {Matarrese}}]{2019MNRAS.484.3818S}
{Sarpa}, E., {Schimd}, C., {Branchini}, E., \& {Matarrese}, S. 2019, \mnras,
  484, 3818, \dodoi{10.1093/mnras/stz278}

\bibitem[{{Schlegel} {et~al.}(2022){Schlegel}, {Kollmeier}, {Aldering},
  {Bailey}, {Baltay}, {Bebek}, {BenZvi}, {Besuner}, {Blanc}, {Bolton},
  {Bonaca}, {Bouri}, {Brooks}, {Buckley-Geer}, {Cai}, {Crane}, {Demina},
  {DeRose}, {Dey}, {Doel}, {Fan}, {Ferraro}, {Finkbeiner}, {Font-Ribera},
  {Gontcho}, {Green}, {Gutierrez}, {Guy}, {Heetderks}, {Huterer}, {Infante},
  {Jelinsky}, {Karagiannis}, {Kent}, {Kim}, {Kneib}, {Kremin}, {Kronig},
  {Konidaris}, {Lahav}, {Lampton}, {Landriau}, {Lang}, {Leauthaud}, {Levi},
  {Liguori}, {Linder}, {Magneville}, {Martini}, {Mateo}, {McDonald}, {Miller},
  {Moustakas}, {Myers}, {Mulchaey}, {Newman}, {Nugent}, {Padmanabhan},
  {Palanque-Delabrouille}, {Piro}, {Poppett}, {Prochaska}, {Pullen},
  {Rabinowitz}, {Raichoor}, {Ramirez}, {Rix}, {Ross}, {Samushia}, {Schaan},
  {Schubnell}, {Seljak}, {Seo}, {Shectman}, {Schlafly}, {Silber}, {Simon},
  {Slepian}, {Slosar}, {Soares-Santos}, {Tarl{\'e}}, {Thompson}, {Valluri},
  {Wechsler}, {White}, {Wilson}, {Y{\`e}che}, {Zaritsky}, \&
  {Zhou}}]{2022arXiv220904322S}
{Schlegel}, D.~J., {Kollmeier}, J.~A., {Aldering}, G., {et~al.} 2022, arXiv
  e-prints, arXiv:2209.04322.
\newblock \doarXiv{2209.04322}

\bibitem[{{Schmidt} {et~al.}(2019){Schmidt}, {Elsner}, {Jasche}, {Nguyen}, \&
  {Lavaux}}]{2019JCAP...01..042S}
{Schmidt}, F., {Elsner}, F., {Jasche}, J., {Nguyen}, N.~M., \& {Lavaux}, G.
  2019, \jcap, 2019, 042, \dodoi{10.1088/1475-7516/2019/01/042}

\bibitem[{{Schmittfull} {et~al.}(2017){Schmittfull}, {Baldauf}, \&
  {Zaldarriaga}}]{2017PhRvD..96b3505S}
{Schmittfull}, M., {Baldauf}, T., \& {Zaldarriaga}, M. 2017, \prd, 96, 023505,
  \dodoi{10.1103/PhysRevD.96.023505}

\bibitem[{{Schmittfull} {et~al.}(2019){Schmittfull}, {Simonovi{\'c}},
  {Assassi}, \& {Zaldarriaga}}]{2019PhRvD.100d3514S}
{Schmittfull}, M., {Simonovi{\'c}}, M., {Assassi}, V., \& {Zaldarriaga}, M.
  2019, \prd, 100, 043514, \dodoi{10.1103/PhysRevD.100.043514}

\bibitem[{{Seljak} {et~al.}(2017){Seljak}, {Aslanyan}, {Feng}, \&
  {Modi}}]{2017JCAP...12..009S}
{Seljak}, U., {Aslanyan}, G., {Feng}, Y., \& {Modi}, C. 2017, \jcap, 2017, 009,
  \dodoi{10.1088/1475-7516/2017/12/009}

\bibitem[{{Seo} {et~al.}(2016){Seo}, {Beutler}, {Ross}, \&
  {Saito}}]{2016MNRAS.460.2453S}
{Seo}, H.-J., {Beutler}, F., {Ross}, A.~J., \& {Saito}, S. 2016, \mnras, 460,
  2453, \dodoi{10.1093/mnras/stw1138}

\bibitem[{{Seo} {et~al.}(2022){Seo}, {Ota}, {Schmittfull}, {Saito}, \&
  {Beutler}}]{2022MNRAS.511.1557S}
{Seo}, H.-J., {Ota}, A., {Schmittfull}, M., {Saito}, S., \& {Beutler}, F. 2022,
  \mnras, 511, 1557, \dodoi{10.1093/mnras/stac082}

\bibitem[{{Shallue} \& {Eisenstein}(2022)}]{2022arXiv220712511S}
{Shallue}, C.~J., \& {Eisenstein}, D.~J. 2022, arXiv e-prints,
  arXiv:2207.12511.
\newblock \doarXiv{2207.12511}

\bibitem[{{Shi} {et~al.}(2018){Shi}, {Cautun}, \& {Li}}]{2018PhRvD..97b3505S}
{Shi}, Y., {Cautun}, M., \& {Li}, B. 2018, \prd, 97, 023505,
  \dodoi{10.1103/PhysRevD.97.023505}

\bibitem[{{Springel}(2005)}]{2005MNRAS.364.1105S}
{Springel}, V. 2005, \mnras, 364, 1105,
  \dodoi{10.1111/j.1365-2966.2005.09655.x10.48550/arXiv.astro-ph/0505010}

\bibitem[{{Takada} {et~al.}(2014){Takada}, {Ellis}, {Chiba}, {Greene},
  {Aihara}, {Arimoto}, {Bundy}, {Cohen}, {Dor{\'e}}, {Graves}, {Gunn},
  {Heckman}, {Hirata}, {Ho}, {Kneib}, {Le F{\`e}vre}, {Lin}, {More},
  {Murayama}, {Nagao}, {Ouchi}, {Seiffert}, {Silverman}, {Sodr{\'e}},
  {Spergel}, {Strauss}, {Sugai}, {Suto}, {Takami}, \&
  {Wyse}}]{2014PASJ...66R...1T}
{Takada}, M., {Ellis}, R.~S., {Chiba}, M., {et~al.} 2014, \pasj, 66, R1,
  \dodoi{10.1093/pasj/pst019}

\bibitem[{{Tegmark}(1997)}]{1997PhRvL..79.3806T}
{Tegmark}, M. 1997, \prl, 79, 3806, \dodoi{10.1103/PhysRevLett.79.3806}

\bibitem[{{Villaescusa-Navarro} {et~al.}(2020){Villaescusa-Navarro}, {Hahn},
  {Massara}, {Banerjee}, {Delgado}, {Ramanah}, {Charnock}, {Giusarma}, {Li},
  {Allys}, {Brochard}, {Uhlemann}, {Chiang}, {He}, {Pisani}, {Obuljen}, {Feng},
  {Castorina}, {Contardo}, {Kreisch}, {Nicola}, {Alsing}, {Scoccimarro},
  {Verde}, {Viel}, {Ho}, {Mallat}, {Wandelt}, \&
  {Spergel}}]{2020ApJS..250....2V}
{Villaescusa-Navarro}, F., {Hahn}, C., {Massara}, E., {et~al.} 2020, \apjs,
  250, 2, \dodoi{10.3847/1538-4365/ab9d82}

\bibitem[{{Villaescusa-Navarro} {et~al.}(2021){Villaescusa-Navarro},
  {Angl{\'e}s-Alc{\'a}zar}, {Genel}, {Spergel}, {Somerville}, {Dave},
  {Pillepich}, {Hernquist}, {Nelson}, {Torrey}, {Narayanan}, {Li}, {Philcox},
  {La Torre}, {Maria Delgado}, {Ho}, {Hassan}, {Burkhart}, {Wadekar},
  {Battaglia}, {Contardo}, \& {Bryan}}]{2021ApJ...915...71V}
{Villaescusa-Navarro}, F., {Angl{\'e}s-Alc{\'a}zar}, D., {Genel}, S., {et~al.}
  2021, \apj, 915, 71, \dodoi{10.3847/1538-4357/abf7ba}

\bibitem[{{Wang} {et~al.}(2014){Wang}, {Mo}, {Yang}, {Jing}, \&
  {Lin}}]{2014ApJ...794...94W}
{Wang}, H., {Mo}, H.~J., {Yang}, X., {Jing}, Y.~P., \& {Lin}, W.~P. 2014, \apj,
  794, 94, \dodoi{10.1088/0004-637X/794/1/94}

\bibitem[{{Wang} \& {Pen}(2019)}]{2019ApJ...870..116W}
{Wang}, X., \& {Pen}, U.-L. 2019, \apj, 870, 116,
  \dodoi{10.3847/1538-4357/aaf231}

\bibitem[{{Wang} {et~al.}(2020){Wang}, {Li}, \& {Cautun}}]{2020MNRAS.497.3451W}
{Wang}, Y., {Li}, B., \& {Cautun}, M. 2020, \mnras, 497, 3451,
  \dodoi{10.1093/mnras/staa2136}

\bibitem[{{Wang} {et~al.}(2022){Wang}, {Zhao}, {Koyama}, {Percival},
  {Takahashi}, {Hikage}, {Gil-Mar{\'\i}n}, {Hahn}, {Zhao}, {Zhang}, {Mu}, {Yu},
  {Zhu}, \& {Ge}}]{2022arXiv220205248W}
{Wang}, Y., {Zhao}, G.-B., {Koyama}, K., {et~al.} 2022, arXiv e-prints,
  arXiv:2202.05248.
\newblock \doarXiv{2202.05248}

\bibitem[{{Wang} {et~al.}(2023){Wang}, {Zhao}, {Zhai}, {Koyama}, {Percival},
  {Guo}, {Li}, {Zhao}, {Nishimichi}, {Gil-Mar{\'\i}n}, {Feng}, {Zhang}, \&
  {Wu}}]{2023arXiv231105848W}
{Wang}, Y., {Zhao}, R., {Zhai}, Z., {et~al.} 2023, arXiv e-prints,
  arXiv:2311.05848, \dodoi{10.48550/arXiv.2311.05848}

\bibitem[{{White}(2015)}]{2015MNRAS.450.3822W}
{White}, M. 2015, \mnras, 450, 3822, \dodoi{10.1093/mnras/stv842}

\bibitem[{{Yu} {et~al.}(2017){Yu}, {Zhu}, \& {Pen}}]{2017ApJ...847..110Y}
{Yu}, Y., {Zhu}, H.-M., \& {Pen}, U.-L. 2017, \apj, 847, 110,
  \dodoi{10.3847/1538-4357/aa89e7}

\bibitem[{{Zhu} {et~al.}(2018){Zhu}, {Yu}, \& {Pen}}]{2018PhRvD..97d3502Z}
{Zhu}, H.-M., {Yu}, Y., \& {Pen}, U.-L. 2018, \prd, 97, 043502,
  \dodoi{10.1103/PhysRevD.97.043502}

\bibitem[{{Zhu} {et~al.}(2017){Zhu}, {Yu}, {Pen}, {Chen}, \&
  {Yu}}]{2017PhRvD..96l3502Z}
{Zhu}, H.-M., {Yu}, Y., {Pen}, U.-L., {Chen}, X., \& {Yu}, H.-R. 2017, \prd,
  96, 123502, \dodoi{10.1103/PhysRevD.96.123502}

\end{thebibliography}
\bibliographystyle{aasjournal}

\end{document}